\documentclass[12pt]{article}
\usepackage[margin=1in]{geometry}
\usepackage{setspace}
\usepackage[utf8]{inputenc} 
\usepackage{amsmath}
\usepackage{natbib}
\usepackage{amssymb}
\usepackage{hyperref}
\usepackage{booktabs}
\usepackage{graphicx}
\usepackage{subcaption}
\usepackage{xr}
\usepackage{mathtools}
\DeclareMathOperator*{\argmin}{arg\,min}
\usepackage{tocloft}
\usepackage{multirow}
\usepackage{float}
\usepackage{bibunits}
\usepackage{bm}
\usepackage{amsthm}
\usepackage{cleveref}
\defaultbibliographystyle{plainnat}
\newtheorem{theorem}{Theorem}
\newtheorem{lemma}{Lemma}
\newtheorem{corollary}{Corollary}
\theoremstyle{definition}
\newtheorem{condition}{Condition}
\theoremstyle{plain}
\newenvironment{keywords}{\vspace{1em}\noindent\textbf{Keywords: }}{\par}
\newcommand{\Hline}{\hline}
\newcommand{\backmatter}{}

\newcommand{\bX}{\mathbf{X}}
\newcommand{\bW}{\mathbf{W}}

\newcommand{\bV}{\mathbf{V}}

\newcommand{\btheta}{\bm{\theta}}
\newcommand{\bbeta}{\bm{\beta}}

\newcommand{\Msc}{\mathcal{M}}

\newcommand{\Bsc}{\mathcal{B}}
\newcommand{\Dsc}{\mathcal{D}}

\newcommand{\mhat}{\widehat{m}}
\newcommand{\fhat}{\hat{f}}
\newcommand{\bthetahat}{\widehat{\btheta}}

\newcommand{\Deltahat}{\hat{\Delta}}

\newcommand{\muhat}{\hat{\mu}}
\newcommand{\Mschat}{\widehat{\Msc}}

\newcommand{\mutilde}{\tilde{\mu}}

\newcommand{\Zbar}{\overline{Z}}

\newcommand{\E}{\mathbb{E}}
\newcommand{\inp}{\overset{p}{\to}}
\newcommand{\bthetabar}{\overline{\btheta}}
\def\transpose{{\sf \scriptscriptstyle{T}}}
\def\trans{^{\transpose}}
\newcommand{\indep}{\perp \!\!\! \perp}
\newcommand{\tpr}{\textrm{TPR}}
\newcommand{\fpr}{\textrm{FPR}}

\newcommand{\acc}{\textrm{ACC}}

\newcommand{\fone}{\textrm{F}1}
\newcommand{\npv}{\textrm{NPV}}
\newcommand{\ppv}{\textrm{PPV}}

\newcommand{\train}{\textrm{train}}

\newcommand{\rhoainv}{\rho_a^{-1}}
\newcommand{\supv}{\text{\tiny{SUP}}}
\newcommand{\ssv}{\text{\tiny{SS}}}

\usepackage{titlesec}

\titlespacing*{\section}
  {0pt}{1.5ex plus 1ex minus .2ex}{1ex plus .2ex}
\titlespacing*{\subsection}
  {0pt}{1ex plus 0.5ex minus .2ex}{0.5ex plus .2ex}
\titlespacing*{\subsubsection}
  {0pt}{1ex plus 0.5ex minus .2ex}{0.5ex plus .2ex}

\usepackage{needspace}

\usepackage{xcolor}
\newcommand{\blue}[1]{\textcolor{black}{#1}}

\title{Reliable fairness auditing with semi-supervised inference}

\author{Jianhui Gao and
Jessica Gronsbell\thanks{Email: \texttt{j.gronsbell@utoronto.ca}} \\
Department of Statistics, University of Toronto, Toronto, ON, Canada}

\begin{document}
\begin{bibunit}









\label{firstpage}

\maketitle

\begin{abstract}
Machine learning (ML) models often exhibit bias that can exacerbate inequities in biomedical applications. Fairness auditing, the process of evaluating a model's performance across subpopulations, is critical for identifying and mitigating these biases. However, audits typically rely on large volumes of labeled data, which are costly and labor-intensive to obtain. To address this challenge, we introduce {\it{Infairness}}, a unified framework for auditing a wide range of fairness criteria using semi-supervised inference. Our approach combines a small labeled dataset with a large unlabeled dataset by imputing missing outcomes via regression with carefully selected nonlinear basis functions. \blue{Through extensive theoretical and empirical analyses, we show that our proposed estimator is (i) robust to specification of the ML or imputation model and (ii) substantially more efficient than supervised estimation based solely on the labeled data. In two real-world fairness audits using electronic health record and medical imaging data, Infairness reduces variance by approximately 50\% compared to supervised estimation, underscoring its value for reliable fairness auditing with limited labeled data.}\\  
\end{abstract}

%





%

\section{Introduction}
\label{s:intro}
Machine learning (ML) is used across healthcare and biomedical research to improve patient outcomes, enhance diagnostic accuracy, and optimize treatment effectiveness \citep{rajpurkar_ai_2022}. However, mounting evidence demonstrates that ML models can exhibit {\it{unfairness}} \blue{by making less favorable decisions for certain groups or individuals \citep{mehrabi_survey_2021}. In a landmark study, Obermeyer et al.\ illustrated the damaging impact of algorithmic bias in a commercial system used to guide treatment decisions for millions of patients in the U.S.\ \citep{obermeyer_dissecting_2019}. The algorithm predicted healthcare cost as a proxy for healthcare need. As a result, Black patients were disproportionately under-identified for high-risk care management programs as they incur lower costs than White patients with comparable medical burden due to structural disparities in healthcare access and treatment. This study fortified a more socially conscious approach to ML deployment and led to further scrutiny of ostensibly accurate models in biomedical applications. A subsequent systematic audit of computer vision algorithms across diverse medical imaging modalities found that nearly all models exhibited performance disparities across protected attributes, including age, sex, and race} \citep{xu_addressing_2024}. A benchmarking study of widely used electronic health record (EHR) phenotyping algorithms for pneumonia and sepsis identified significant variability in diagnostic performance across groups defined by gender, race, and ethnicity \citep{ding_identify_2024}. \blue{These findings collectively highlight that algorithmic unfairness is not only prevalent, but that it can perpetuate and deepen existing inequity if left unaddressed.} 

\blue{In response, researchers now utilize a variety of statistical methods to discover and quantify such disparities, a process referred to as ``group fairness auditing'' \citep{rajkomar_ensuring_2018}. While a central component of responsibly deploying ML systems, reliable audits require sufficiently large labeled datasets to detect differences in subgroup performance. This is a major bottleneck in biomedical applications as labeled data is often scarce due to the specialized expertise, time, and cost required for annotation}. For instance, it takes an annotator 2-15 minutes to label a medical image whereas commonplace images, such as those in ImageNet, can be labeled at a rate of two images per second through crowdsourcing \citep{li_crowdsourcing_2010}. In the analysis of EHR data, labeling a single discharge summary can take up to 10 minutes and often involves multiple experienced clinicians to ensure consensus \citep{gehrmann_comparing_2018}.

\blue{Consequently, fairness audits within the biomedical domain often fall within a semi-supervised (SS) setting, in which a small labeled dataset is accompanied by a much larger unlabeled one. Traditional supervised auditing approaches rely solely on the labeled data, which limits estimation precision and increases the risk of missing performance disparities. While SS inference has been widely used to improve efficiency by leveraging both labeled and unlabeled data, it has yet to be rigorously developed in the context of group fairness auditing. To fill this gap, we introduce a unified SS framework, which we term {\emph{Infairness}}, that improves the precision of group fairness audits and, in turn, enables more powerful detection of subgroup differences without increasing labeling demands.}

The remainder of this article is organized as follows. \blue{Section \ref{s:rw} reviews relevant literature.} Section \ref{s:pre} introduces the problem set-up. Section \ref{s:methods} details the methodology. Section \ref{s:asymp} presents theoretical results. \blue{Sections \ref{s:sim} and \ref{s:real} presents simulated and real data examples, respectively.} We conclude in Section \ref{s:diss} with a discussion of the implications of our work.

\section{\blue{Related literature}}\label{s:rw}
\blue{We first review related areas of statistics and ML to situate our proposal within the context of classical and modern literature.} 

\subsection{Missing data and semi-supervised (SS) inference}
\blue{The goal of SS inference is to determine whether and how unlabeled data can be leveraged to improve the efficiency of supervised estimators based on the labeled data. SS inference is a special case of a missing data problem in which the positivity assumption is violated due to the disparate sizes of the labeled and unlabeled data. Existing work has focused on familiar problems such as mean estimation, regression, and M-estimation \citep[e.g.,][]{zhang_semi-supervised_2019,song_general_2023, xu2025unified}. An underexplored area is model performance evaluation. \citet{gronsbell_semi-supervised_2018} proposed a nonparametric method for estimation of the ROC curve and \citet{gronsbell_efficient_2022} extended these ideas for estimation of the misclassification error and calibration. Related approaches have also appeared within transfer and weakly supervised learning \citep[e.g.,][]{gao_semi-supervised_2024, kiyasseh_framework_2024}. These methods, however, are designed for overall model performance evaluation rather than fairness auditing.}

\subsection{Prediction-powered inference}
\blue{SS inference is closely related to prediction-powered inference (PPI). PPI is a modern extension of a missing problem, where predictions from pre-trained ML models are used to impute missing outcomes \citep{doi:10.1126/science.adi6000}. PPI has been primarily studied for Z-estimation and only recently extended for evaluating binary classification performance \citep{pmlr-v267-boyeau25a}. PPI approaches reduce the variance of supervised estimation with a control-variate derived from the predictions in the unlabeled data. Our proposal instead uses an imputation-based approach and differs in two important ways. First, PPI methods focus on asymptotic regimes where the labeled size grows proportionally with the total sample size. Many biomedical settings fall within the SS framework in which positivity is violated, which our method accommodates. Second, we propose to recalibrate the predictions using auxiliary covariates that are often available during fairness auditing to further improve efficiency.}

\subsection{Group fairness methods}
\blue{Traditionally, group fairness audits rely on supervised estimation based on labeled data, which can lead to high variance when subgroup sample sizes are small. Several methods address this limitation by borrowing information from the full labeled dataset. \citet{miller_model-based_2021} proposed model-based estimators for subgroup accuracy measures while \citet{van_breugel_can_2023} used generative models to augment labeled data with synthetic data. These approaches, however, do not leverage unlabeled data. The only existing SS method for fairness auditing is the beta calibration (BC) approach of \cite{ji_can_2020}, which targets a small number of fairness metrics. While flexible, BC relies on parametric assumptions that, when violated, can lead to biased, albeit very precise, inferences about subgroup performance. This behavior can foster overconfidence in a distorted understanding of a model's performance and undermine the scientific rationale for utilizing SS inference. Our proposal is guaranteed to provide consistent estimation and valid inference without relying on parametric assumptions.}

\section{Preliminaries}\label{s:pre}
\subsection{Problem set-up}\label{s:set-up}
We consider an ML model for predicting a binary outcome $Y$. As our focus is on group fairness auditing, we aim to evaluate whether the performance of the model differs across groups defined by a categorical protected attribute $A$. The ML model is trained on features $\bX$, which may or may not include $A$. $\bX$ need not be a vector and may be an image, text, tensor, or other type of data. During validation, a vector of auxiliary covariates $\bW = (W_1, \dots, W_p)\trans$ if often available for analysis. For example, in medical imaging or EHR applications, $Y$ may be a disease of interest, $\bX$ a chest X-ray or discharge summary, and $\bW$ a set of additional protected attributes or clinical variables not utilized during model training.

With a slight abuse of notation, the validation data therefore consists of a {\it{labeled}} dataset with $n$ independent and identically distributed (i.i.d.) samples, $\{(Y_i, \bX_i\trans, \bW_i\trans, A_i)\trans\mid i = 1, ..., n\}$, together with an independent {\it{unlabeled}} dataset with $N$ i.i.d. realizations of \newline $\{( \bX_{i}\trans, \bW_{i}\trans, A_{i})\trans\mid i = n+1, ..., n+N\}$. We focus on the classical SS setting in which it is assumed that (i) the labeled and unlabeled data arise from the same underlying distribution (or equivalently that $Y$ is missing completely at random) and (ii) $n/N \to 0 $ as $n \to\infty$ \cite[e.g.,][]{chakrabortty_efficient_2018, zhang_semi-supervised_2019}. This setting occurs when an investigator has access to a large set of unlabeled data and a limited budget is available to select samples uniformly at random for labeling such as in the imaging and EHR examples from Section \ref{s:intro}. 

To ensure that group-specific performance metrics can be estimated, we further assume that there is sufficient labeled and unlabeled data within each of the groups defined by $A$. To this end, let $n_a = \sum_{i=1}^n I(A_i = a)$ and $N_a = \sum_{i=n+1}^{n+N} I(A_i = a)$ denote the number of labeled and unlabeled observations in the group with $A = a$, respectively. We assume that $n_a/n \to \rho_a \in (0,1)$  as $n \to\infty$ and $N_a/N \to \rho_a \in (0,1)$  as $N \to\infty$ for each $a$. These assumptions, together with assumption (ii), imply that $n_a/N_a \to 0 $ as $n \to\infty$.

We make no assumptions on the ML model, denoted as $\fhat$,  other than that it is trained on a dataset, $\Dsc_{\train}$, that is independent of the validation dataset. In practice, models are commonly trained with weakly-supervised methods that accommodate scenarios with scarce labeled data or simply use pre-trained models, such as LLMs or computer vision algorithms \cite[e.g.,][]{yu_enabling_2018, seyyed-kalantari_underdiagnosis_2021}. The predictions of $Y$ are denoted as $S = \fhat(\bX)$ and are available for all $(n+N)$ observations in the validation data. The final classification for $Y$ is $D = I(S\ge c)$ for some pre-specified threshold $c \in (0, 1)$. 

\subsection{Parameter of interest}
To quantify the performance of the ML model in predicting $Y$ within groups defined by $A$, we consider a wide range of group fairness metrics based on accuracy, discrimination, and calibration. Table \ref{tab:metrics} summarizes common group-specific metrics that measure model performance within each group with $A = a$, denoted generally as $\Msc_a$. For example, the group-specific TPR is defined as $\tpr_a \coloneq {P}\left(D = 1 \mid Y = 1, A = a\right).$ Note that the probability is taken over the distribution of $(Y, D, A)$ conditioning on  $\mathcal{D}_{\textrm{train}}$. To simplify notation, we omit explicit conditioning on $\mathcal{D}_{\textrm{train}}$, with the understanding that all target parameters and corresponding inference procedures are defined conditionally on the training data. To unify notation for the various group-specific metrics, we define $\mu_a^Z = \E(Z \mid A = a)$
for a random variable $Z$. For example, the group-specific TPR is
$\tpr_a = \frac{\mu_a^{DY}}{\mu_a^Y}.$ 

Existing work within the fairness literature primarily focuses on comparing group-specific metrics across two groups defined by a binary protected attribute \citep{mehrabi_survey_2021, gao_what_2024}. We similarly focus on the two-group setting and discuss extensions to the multi-group setting in Section \ref{s:diss}. In the two-group setting, fairness criteria are most often defined in terms of the difference between the group-specific metrics, denoted as $\Delta_{\Msc} = \Msc_0 - \Msc_1$ and the goal is to conduct inference on $\Delta_{\Msc}$. Table \ref{tab:criteria} summarizes common group fairness criteria \citep{hardt_equality_2016}. Again using the TPR as an example, $\Delta_{\tpr} = \tpr_0 - \tpr_1$, and when $\Delta_{\tpr} = 0$ the model is said to satisfy the {\it{equal opportunity}} criterion. For exposition, we use equal opportunity as a running example in Section \ref{s:methods}.

\begin{table}[htbp!]
\caption{\textbf{Definitions and unified notation for group-specific performance metrics.} Notations: $Y$: outcome, $A$: protected attribute, $S$: model prediction, $D$: model classification based on thresholding $S$, $\Msc_a$: model performance metric within the group with $A = a$, $\mu_a^Z = \E(Z\mid A = a)$. Acronyms: TPR: True positive rate, FPR: False positive rate, PPV: Positive predictive value, NPV: Negative predictive value, , $F1$: F$_1$-score, ACC: Accuracy, BS: Brier score.}
\label{tab:metrics}
\begin{center}
\begin{tabular}{lcc}
\Hline
$\Msc_a$ & Definition & Unified Notation  \\ \hline
   $\textrm{TPR}_a$ & $P\left(D = 1 \mid Y=1, A = a\right)$ & $\mu_a^{DY}/\mu_a^Y$    \\
   $\textrm{FPR}_a$  & $P\left(D = 1 \mid Y=0, A = a \right)$ & $(\mu_a^D - \mu_a^{DY})/(1- \mu_a^Y)$\\
   $\textrm{PPV}_a$  & $P\left(Y = 1 \mid D = 1, A = a \right)$ &  $\mu_a^{DY}/\mu_a^D$ \\
   $\textrm{NPV}_a$  & $P\left(Y = 0 \mid D = 0, A = a\right)$ & $(1 - \mu_a^D - \mu_a^Y + \mu_a^{DY})/(1 - \mu_a^D)$ \\
   $F1_{a}$ & $2/\left(\textrm{TPR}_a^{-1} + \textrm{PPV}_a^{-1}\right)$ & $2\mu_a^{DY}/(\mu_a^D+\mu_a^Y)$ \\
   $\textrm{ACC}_a$  & $1 - \E\left\{\left(Y - D\right)^2 \mid A =a \right\}$ & $1 - \mu_a^Y - \mu_a^D + 2 \mu_a^{DY}$ \\
   $\textrm{BS}_a$  & $\E\left\{\left(Y-S\right)^2 \mid A = a \right\}$ & $\mu_a^{S^2} - 2\mu_a^{SY} + \mu_a^Y$\\
  \hline
\end{tabular}
\end{center}
\end{table}

\begin{table}[htbp!]
\caption{\textbf{Common group fairness criteria and corresponding definitions in the setting of a binary protected attribute.} A model satisfies a given fairness criterion if the corresponding fairness measure(s), $\Delta_{\Msc} = \Msc_0 - \Msc_1$, is zero. Common choices of group-specific performance, $\Msc_a$, include the true positive rate (TPR), false positive rate (FPR), positive predictive value (PPV), negative predictive value (NPV), accuracy (ACC), Brier score (BS), and F1 score (F1).}
\label{tab:criteria}
\begin{center}
\begin{tabular}{ll}
\Hline
Fairness Criterion & Definition \\ \hline
Equal Opportunity & $\Delta_{\tpr} = 0$\\
Predictive Equality & $\Delta_{\fpr} = 0$ \\
(Positive) Predictive Parity & $\Delta_{\ppv} = 0$ \\
(Negative) Predictive Parity & $\Delta_{\npv} = 0$ \\
F1 Score Parity & $\Delta_{\textrm{F1}} = 0$ \\
Overall Accuracy Equality & $\Delta_{\acc} = 0$\\
Brier Score Parity & $\Delta_{\textrm{BS}} = 0$\\
  \hline
\end{tabular}
\end{center}
\end{table}
\vspace{-2em}

\section{Methods}\label{s:methods}
\subsection{Review of supervised inference for group fairness auditing}
We \blue{first revisit the traditional supervised approach} to evaluating group fairness using only the labeled data. Constructing supervised estimators for the metrics in Table \ref{tab:metrics} requires estimating $\mu_a^Z$ for $Z \in \{Y, D, S^2, SY, DY\}$ within the labeled data as $n_{a}^{-1}\sum_{i=1}^{n} Z_i I(A_i = a)$. Supervised estimators for $\Msc_a$ and $\Delta_{\Msc}$ are then obtained by substituting the estimator of $\mu_{a}^Z$ into the formulas provided in Table \ref{tab:metrics}. For example, the estimator of $\Delta_{\tpr}$ is
\begin{equation*}
   \widehat{\Delta}^{\supv}_{\tpr} \coloneq \widehat{\tpr}^{\supv}_0 - \widehat{\tpr}^{\supv}_1 = \frac{n_0^{-1}\sum_{i=1}^{n} D_iY_i I(A_i = 0)}{n_0^{-1}\sum_{i=1}^{n} Y_i I(A_i = 0)}-\frac{n_1^{-1}\sum_{i=1}^{n} D_iY_i I(A_i = 1)}{n_1^{-1}\sum_{i=1}^{n} Y_i I(A_i = 1)}.
\end{equation*}
To conduct inference on $\Delta_{\tpr}$, large sample confidence intervals can be constructed based on the asymptotic normality of $\widehat{\Delta}^{\supv}_{\tpr}$ detailed in Theorem \ref{thm:sup-if} in the Supplementary Material. 

\subsection{Infairness: Proposed SS inference method for group fairness auditing} \label{sec: infairness}
The key distinction between the SS and supervised approaches to inference is that the SS approach uses the large unlabeled dataset alongside the small labeled dataset to improve estimation efficiency. Intuitively, the efficiency gain arises from exploiting the abundant unlabeled data within each group defined by $A$, which provides near-complete information on the joint distribution of the prediction, $S$, and the auxiliary covariates, $\bW$. \blue{This intuition provides the underlying motivation for our proposal, {\it{Infairness}}.} 

Analogously to the supervised setting, constructing SS estimators for the group fairness metrics in Table \ref{tab:metrics} requires estimation of $\mu_a^Z$ for $Z \in \{Y, D, S^2, SY, DY\}$. The Infairness estimators for $\Msc_a$ and $\Delta_{\Msc}$ are obtained by substituting the SS estimators of $\mu_{a}^Z$ into the formulas provided in Table \ref{tab:metrics}. To this end, we consider two cases. First, consider $\mu_a^{Z}$ for $Z \in \{Y, SY, DY\}$ as these functionals depend on $Y$, which is only available in the labeled data. To illustrate the potential utility of the unlabeled data in estimation, note that $\mu_a^{Z} = \E( Z \mid A = a) $ can be rewritten as 
\begin{equation}\label{eq: muZ-Y}
\mu_a^{Z} = \E \{ \phi(Z) \E( Y  \mid S, \bW, A = a) \mid A =a \}  \quad \mbox{where} \quad  
\phi(Z) = 
\begin{cases}
1 & \text{if } Z = Y \\
S & \text{if } Z = SY \\
D & \text{if } Z = DY.
\end{cases}
\end{equation}
The expression in \eqref{eq: muZ-Y} highlights that $\mu_a^Z$ inherently depends on the distribution of $(S, \bW\trans)\trans \mid A = a$  and hence estimation may be improved by making careful use of the unlabeled data. Moreover, a natural SS estimator can be obtained by constructing the empirical analogue of \eqref{eq: muZ-Y}. Specifically, we estimate $\mu_a^{Z}$ by (i) imputing the missing $Y$ with an estimate of $\E( Y  \mid S, \bW,  A = a)$ learned from the labeled data and (ii) averaging the resulting imputations over the unlabeled data in the group with $A = a$. The potential efficiency gain relative to supervised estimation hinges on the quality of the imputation in step (i) and is driven by step (ii), which leverages the abundant unlabeled data. In contrast, for the second case with  $Z \in \{D, S^2\}$, $\mu_a^{Z}$ does not depend on $Y$ and therefore estimation does not require imputation. We can simply obtain estimators for $\mu_a^{Z}$ with the corresponding empirical averages in the unlabeled data. Similar to the imputation-based SS estimators, these estimators offer improved precision relative to their supervised counterparts due to averaging over a much larger sample size. 

\subsubsection{Imputation strategy}
\blue{The key challenge in an imputation-based approach is balancing robustness with efficiency. The imputations must be constructed in such a way that the Infairness estimator is consistent for its target while also yielding precision gains from the unlabeled data. A fully nonparametric method achieves both goals, but generally does not perform well when the dimension of $(S, \bW\trans)\trans$ is moderate due to the curse of dimensionality \citep{chakrabortty_efficient_2018, tan2025efficient}. We therefore propose to impute $Y$ within each group by fitting the \textit{working} model} 
\begin{equation}\label{eq:imputation}
     \E( Y  \mid S, \bW,  A = a) = g\left(\btheta_{a}\trans \mathcal{B}_a \right) 
\end{equation}
where $\mathcal{B}_a = \mathcal{B}_a(S, \bW)$ is a set of basis functions of fixed dimension and $g(\cdot):\mathbb{R}\to [0, 1]$ is a specified smooth monotone function such as the expit function. \blue{The basis expansion can include nonlinear and interaction effects to achieve a flexible representation of $\E( Y  \mid S, \bW,  A = a)$ and can also be carefully designed to ensure consistency of the Infairness estimator. With respect to the former, we suggest using flexible spline models and study the practical impact of the choice of basis in our empirical studies in Section \ref{s:sim}. With respect to the latter,} we propose to estimate $\btheta_{a}$ from the labeled data with $\bthetahat_{a}$, which is the solution to
\begin{equation}
\label{eq: theta}
n_a^{-1}\sum_{i=1}^n \mathcal{B}_{a,i}\left\{Y_i - g\left(\btheta_{a}\trans \mathcal{B}_{a,i}\right)\right\}I(A_i = a) - \lambda_{n_a}\btheta_{a} = \bm{0}
\end{equation}
and $\lambda_{n_a}= o\left(n_a^{-\frac{1}{2}}\right)$ is a penalty term used to stabilize model fitting. We show in Supplementary Section \ref{supp:thetahata} that $\bthetahat_a$ is consistent for $\bthetabar_a$, defined as the solution to $ \E[\Bsc_{a}\{Y-g(\btheta_a\trans\Bsc_a)\}I(A=a)] = \bm{0}$. The imputations are computed as $\mhat_a = g(\bthetahat_{a}\trans \mathcal{B}_a)$ within the unlabeled data and used to evaluate the fairness criteria. For example, returning to equal opportunity, the Infairness estimator of $\Delta_{\tpr}$ is:
$$ \widehat{\Delta}^{\ssv}_{\tpr} = \widehat{\tpr}^{\ssv}_0 - \widehat{\tpr}^{\ssv}_1 = \frac{N_0^{-1}\sum_{i=n+1}^{n+N}D_i\mhat_{0,i} I(A_i = 0)}{N_0^{-1}\sum_{i=n+1}^{n+N} \mhat_{0,i}  I(A_i = 0)} - \frac{N_1^{-1}\sum_{i=n+1}^{n+N} D_i\mhat_{1,i}  I(A_i = 1)}{N_1^{-1}\sum_{i=n+1}^{n+N} \mhat_{1,i}  I(A_i = 1)}.$$  

Importantly, the Infairness estimators are consistent for their targets provided that SS estimators of $\mu_a^Z$ are consistent for their targets. For $Z\in\{Y,SY, DY\}$, $\mu_a^Z$ depends on $Y$ and the SS estimators are based on the imputations for $Y$. We therefore require that
\begin{equation}\label{eq:suff_cond}
\begin{aligned}
&\E\{ (1, D, S)\trans (Y - \mhat_a)  I(A = a) \mid \mhat_a\} \inp {\bf{0}} \mbox{ as } n_a \to \infty .\\   
\end{aligned}
\end{equation}
\blue{While this condition is satisfied if the working model in \eqref{eq:imputation} is correctly specified, the functional form of $\E( Y  \mid S, \bW,  A = a)$ may be misspecified in practice. We can, however, ensure that the condition \eqref{eq:suff_cond} is satisfied by simply including an intercept, $S$, and $D$ in the basis functions as $\bthetahat_a$ is derived from the estimating equation in \eqref{eq: theta}.}

\section{Asymptotic results}\label{s:asymp}
We next summarize the key asymptotic properties of the Infairness estimator in Theorem \ref{thm:ss-if}. Detailed derivations are provided in Sections \ref{supp:sup} -- \ref{supp:ss}. 

\begin{theorem}[\textbf{Influence function of $\Deltahat^{\ssv}_{\Msc}$}]\label{thm:ss-if}
Under the assumptions in Section \ref{s:set-up} and regularity conditions in Section \ref{supp:ass}, $\Deltahat^{\ssv}_{\Msc}\inp\Delta_{\Msc}$ and
\begin{equation*}
 n^{\frac{1}{2}} \left(\Deltahat^{\ssv}_{\Msc} - \Delta_{\Msc}\right)  =  n^{-\frac{1}{2}} \sum_{a \in \{0, 1\}} (-1)^{a}\rho_a^{-1}\sum_{i=1}^n \textnormal{\bf IF}^{\ssv}_{\Msc_a}\left(Y_i, \Bsc_{a,i};c, \bthetabar_a\right)I(A_i=a) +o_p(1)
\end{equation*}
where $\textnormal{\bf IF}^{\ssv}_{\Msc_a}(Y_i, \Bsc_{a,i};c, \bthetabar_a)$ is specified in Table \ref{tab:if}. Therefore, $n^{\frac{1}{2}}\left(\Deltahat^{\ssv}_{\Msc} - \Delta_{\Msc}\right)$ converges weakly to a zero-mean Gaussian distribution with variance $$\sum_{a\in\{0,1\}}\rho_a^{-1}\E\left[\left\{\textnormal{\bf IF}^{\ssv}_{\Msc_a}\left(Y_i, \Bsc_{a,i};c, \bthetabar_a\right)\right\}^2\mid A = a\right].$$
\end{theorem}

\begin{table}[h]
\caption{\textbf{Influence functions of $\Deltahat^{\supv}_{\Msc}$ and $\Deltahat^{\ssv}_{\Msc}$.} Notations: $Y$: outcome, $S$: model prediction, $c$: cutoff, $D$: model classification based on thresholding $S$ at $c$, $\Msc_a$: model performance metric within the group with $A = a$, $\mu_a^Z = \E(Z\mid A = a)$, $\Bsc_a = \Bsc_a(S, \bW)$: a finite set of basis functions of fixed dimension that includes an intercept, $g(\cdot): \mathbb{R}\to[0,1]$: a specified smooth monot
one function, $\bthetabar_a$: solution to $\E[\Bsc_a\{Y-g(\btheta\trans\Bsc_a)\}I(A=a)] = 0$. Acronyms: TPR: True positive rate, FPR: False positive rate, PPV: Positive predictive value, NPV: Negative predictive value, $F1$: F$_1$-score, ACC: Accuracy, BS: Brier score. }
\label{tab:if}
\centering
\resizebox{\linewidth}{!}{
    \begin{tabular}{lll}
    \Hline
        $\Msc_a$ &  $\textnormal{\bf IF}^{\supv}_{\Msc_a}(Y_i, S_i;c)$ & $\textnormal{\bf IF}^{\ssv}_{\Msc_a}(Y_i, \Bsc_{a}(S_i,\bW_i);\bthetabar_a, c)$\\
        \hline
        $\tpr_a$ & $(\mu_a^Y)^{-1}\{Y_i(D_i - \tpr_a)\}$ & $(\mu_a^Y)^{-1}\{Y_i- g(\bthetabar_a\trans\Bsc_{a,i})\}\left(D_i - \tpr_a\right)$\\
        $\textrm{FPR}_a$ &$(1-\mu_a^Y)^{-1} \left\{(1-Y_i)(D_i - \textrm{FPR}_a )\right\}$ & $(1-\mu_a^Y)^{-1} \{Y_i- g(\bthetabar_a\trans\Bsc_{a,i})\}\left(\textrm{FPR}_a - D_i\right)$\\
        $\textrm{PPV}_a$ & $ (\mu_a^D)^{-1} \left\{D_i(Y_i - \textrm{PPV}_a)\right\}$ & $(\mu_a^D)^{-1}D_i\{Y_i- g(\bthetabar_a\trans\Bsc_{a,i})\}$ \\
        $\textrm{NPV}_a$ & $ (1-\mu_a^D)^{-1} \left\{(1-D_i)(1 - Y_i - \textrm{NPV}_a)\right\}$ & $(1-\mu_a^D)^{-1}(D_i-1)\{Y_i- g(\bthetabar_a\trans\Bsc_{a,i})\}$\\
        $\textrm{F1}_a$ & $ \left(\mu_a^D + \mu_a^Y\right)^{-1}\left\{D_i\left(Y_i - \textrm{F1}_a\right) + Y_i \left(D_i - \textrm{F1}_a\right)\right\}$ & $\left(\mu_a^D + \mu_a^Y\right)^{-1}\{Y_i- g(\bthetabar_a\trans\Bsc_{a,i})\}\left(2D_i - F1_a\right)$ \\
        $\textrm{ACC}_a$ & $1-(Y_i-D_i)^2 - \textrm{ACC}_a$ & $\{Y_i- g(\bthetabar_a\trans\Bsc_{a,i})\}(2D_i - 1)$\\
        $\textrm{BS}_a$ & $(S_i-Y_i)^2-\textrm{BS}_a$ & $\{Y_i- g(\bthetabar_a\trans\Bsc_{a,i})\}(1-2S_i)$\\
        \hline
    \end{tabular}
}
\end{table}
Theorem \ref{thm:sup-if} in the Supplementary Materials provides similar results for the supervised estimator. We utilize Theorems \ref{thm:sup-if} and \ref{thm:ss-if} to compare the asymptotic variances of the two estimators in the following corollary, with the proof detailed in Section \ref{supp:var-comp}.

\begin{corollary}[\textbf{Variance comparison of $\Deltahat^{\supv}_{\Msc}$ and $\Deltahat^{\ssv}_{\Msc}$}]\label{cor:var-comp}
When the imputation model in equation \eqref{eq:imputation} is correctly specified and $\E(Y\mid S, \bW, A = a) \ne \E(Y \mid A = a)$, 
$$
\sum_{a\in\{0,1\}}\rho_a^{-1}\E\left[\left\{\textnormal{\bf IF}^{\ssv}_{\Msc_a}(Y_i, \Bsc_{a,i};c, \bthetabar_a)\right\}^2\mid A = a\right] < \sum_{a\in\{0,1\}}\rho_a^{-1}\E\left[\left\{\textnormal{\bf IF}^{\supv}_{\Msc_a}(Y_i, S_i;c)\right\}^2\mid A = a\right].
$$
\end{corollary}
Corollary \ref{cor:var-comp} provides a sufficient condition under which the Infairness estimator is theoretically guaranteed to be more efficient than the supervised estimator. Specifically, this occurs when the group-specific imputation models are correctly specified and when the unlabeled data contains meaningful information about $Y$. This result underlies our motivation in constructing the Infairness estimator, that is, the imputation model should be specified flexibly enough so that it can extract as much information as possible from the unlabeled data. While we cannot provide strict theoretical guarantees when the imputation model is misspecified, our proposal will generally yield efficiency gains provided the imputation model is a close approximation to $\E(Y\mid S, \bW, A = a)$. We corroborate this heuristic justification in our empirical studies in Sections \ref{s:sim} and \ref{s:real}. 

\section{\blue{Simulation studies}}\label{s:sim}
\blue{We consider two sets of simulation studies. First, we consider a stylized setting in which we directly generate $S$ to provide intuition for the Infairness procedure and to illustrate its potential benefit over its comparators. Second, we consider a more traditional set-up where we fit an ML model on an independent dataset to generate $S$. For both studies, we present results for $n = 400$ and $N = 20,000$ and include results for $n = 1000$ in Section \ref{supp:sim-large}.}

\subsection{\blue{Evaluation metrics and methods}}
\blue{We compare the supervised estimator with 3 SS approaches: the beta calibration (BC) approach of \cite{ji_can_2020}, the nonparametric (NP) approach of \cite{gronsbell_semi-supervised_2018}, and the proposed Infairness method. The BC and NP approaches base imputation entirely on $S$ and do not utilize $\bW$. The BC approach imputes the missing $Y$ with beta calibrated predictions while the NP approach estimates $\E(Y \mid S)$ with kernel smoothing. For Infairness, we impute $Y$ using natural cubic spline regression models with three knots based on (i) $S$ alone (Infairness($S$)), (ii) $S$ with additive effects of $\bW$ (Infairness($S + \bW$)), and (iii) $S$ with interaction effects between $S$ and $\bW$ (Infairness($S \times \bW$)). Further details regarding the implementation of each method are provided in Section \ref{supp:tuning}.}

\blue{We assess bias, coverage probability (CP), and relative efficiency (RE). Bias is evaluated by comparing average point estimates to their true values. The true parameter values are approximated using supervised estimates derived from a fully labeled dataset of size $10^6$. CP is assessed by the empirical coverage of 95\% confidence intervals constructed using asymptotic standard errors. \blue{We utilize 10-fold cross-validation for standard error estimation for all methods to improve finite sample performance. Further details are provided in Section \ref{supp:tuning}.} RE is defined as the ratio of the mean squared error (MSE) of the supervised estimator to that of each SS method. All results are summarized over 10{,}000 Monte Carlo replicates.}

\subsection{\blue{Stylized setting}}
\blue{We consider two scenarios in the stylized setting. In both scenarios, we generate $Y \sim \mathrm{Bernoulli}(0.3)$, $A \sim \mathrm{Bernoulli}(0.4)$, and an auxiliary covariate $W \sim \mathrm{Bernoulli}(0.5)$. The prediction $S$ was generated so that $S \mid Y = y, A =a, W = w \sim \mathrm{Beta}(\alpha_{y, w, a}, \beta_{y, w, a})$. Scenario 1 considers an ideal setting in which $S \perp (A, W) \mid Y$ so that the assumption of the BC method is satisfied and $S$ performs equally well across groups defined by $A$ and $W$. Scenario 2 considers a setting in which the distribution of $S \mid Y$ also depends on $(A, W)$ and $S$ has variable performance across groups. In both scenarios, $D = I(S > 0.5)$. A detailed description of the data generation is provided in Supplemental Section \ref{supp:sim-dgp-stylized}.}  

\begin{figure}
    \centering
    \includegraphics[width=\linewidth]{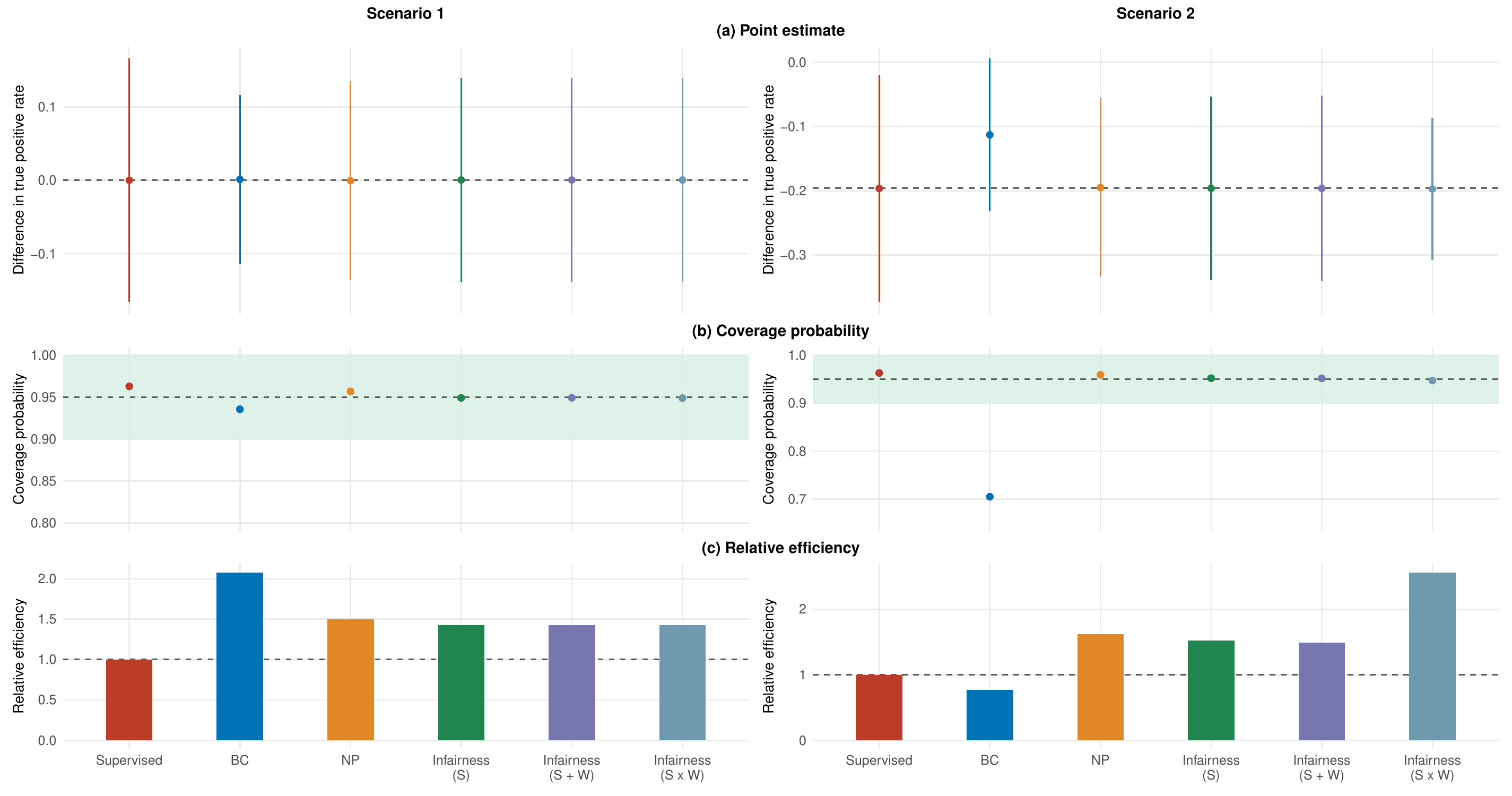}
    \caption{\blue{\textbf{Stylized setting for estimating the difference in true positive rates.} Results are based on $10^4$ Monte Carlo replicates with 20,000 validation observations and 400 labeled observations per replicate. Panel (a) shows the average point estimate of \(\Delta_{\tpr}=\tpr_0-\tpr_1\), with vertical lines corresponding to the average estimate plus or minus 1.96 times the empirical standard error; the dashed horizontal line marks the truth. Panel (b) shows empirical coverage probability of nominal 95\% confidence intervals, with the shaded region indicating coverage between 0.90 and 1.00. Panel (c) shows relative efficiency, defined as the ratio of the supervised mean squared error to the mean squared error of each method. }} \label{fig:toy}
\end{figure}

\blue{For simplicity, we continue with our running example of equal opportunity. Figure \ref{fig:toy} summarizes the results. In Scenario 1, BC, NP, and the Infairness estimators exhibit minimal bias, achieve nominal coverage, and improve efficiency relative to supervised estimation. The BC method is the most efficient approach as its parametric assumptions are satisfied. However, in Scenario 2, the assumptions of the BC method are violated, which results in coverage near 70\% and substantial underestimation of the difference in the group-specific TPRs. The BC method also yields estimates with improved efficiency over supervised estimation. Practically, this translates into an overconfidence that $S$ is less biased than it truly is. With $n = 1000$, CP deviates further from the nominal level and falls below 40\% (Figure~\ref{fig:toy-1000}). The NP method remains approximately unbiased with near-nominal coverage, but is less efficient than Infairness($S\times W$), which effectively leverages the additional information in $W$. As the additive effect of $W$ is minimal by design in this scenario, the alternative Infairness approaches do not properly utilize the information in $W$. However, both approaches do not incur efficiency loss relative to NP. Although an oversimplification of a real-world audit, these findings underscore the gap that Infairness fills: it always provides consistent estimation of the target parameter while also improving over NP when additional information in $W$ is adequately incorporated into the imputation model. This take-away is reinforced in the more complex simulation scenarios considered in Section \ref{sim: trad-setting}.}

\subsection{\blue{Traditional setting}}\label{sim: trad-setting}
\blue{We next generate $(\bX^\top, \bW^\top, A)^\top$ from a 16-dimensional mean-zero multivariate normal. The covariance between the $k$th and $\ell$th random variables is $3 \cdot (0.4)^{|k-\ell|}$. The first 10 variables are used as features $\bX$ for model training, the next 5 as auxiliary covariates $\bW$ available at auditing, and the final variable is thresholded to form the binary protected attribute $A$ with prevalence 0.4. The outcome $Y$ is generated under four scenarios of varied complexity that impact the fit of the ML model used to generate $S$. Scenario A uses a logistic model in which the linear predictor contains nonlinear and interaction effects. Scenario B applies a nonlinear transformation to the linear predictor. Scenario C uses a complementary log-log model with a nonlinear transformation of the linear predictor. Scenario D uses a tree-based model in which the probability of the outcome is constant within regions defined by $\bX$, $\bW$, and $A$. Detailed data generating mechanisms are provided in Section \ref{supp:sim-dgp-traidtional}. In Section \ref{supp:uninformative}, we also consider an extreme scenario in which $S$ is completely uninformative of $Y$. In all scenarios, $S$ is obtained by fitting a logistic regression model of $Y$ on $\bX$ using an independent labeled dataset of size $10^4$ generated from the same distribution. The final classification is taken as $D = I(S > 0.5)$. While selecting appropriate metrics for fairness auditing is context-dependent, we consider all metrics in Table \ref{tab:criteria} for the purposes of illustration.}

\begin{figure}[htbp!]
    \centering
    \includegraphics[width=\linewidth]{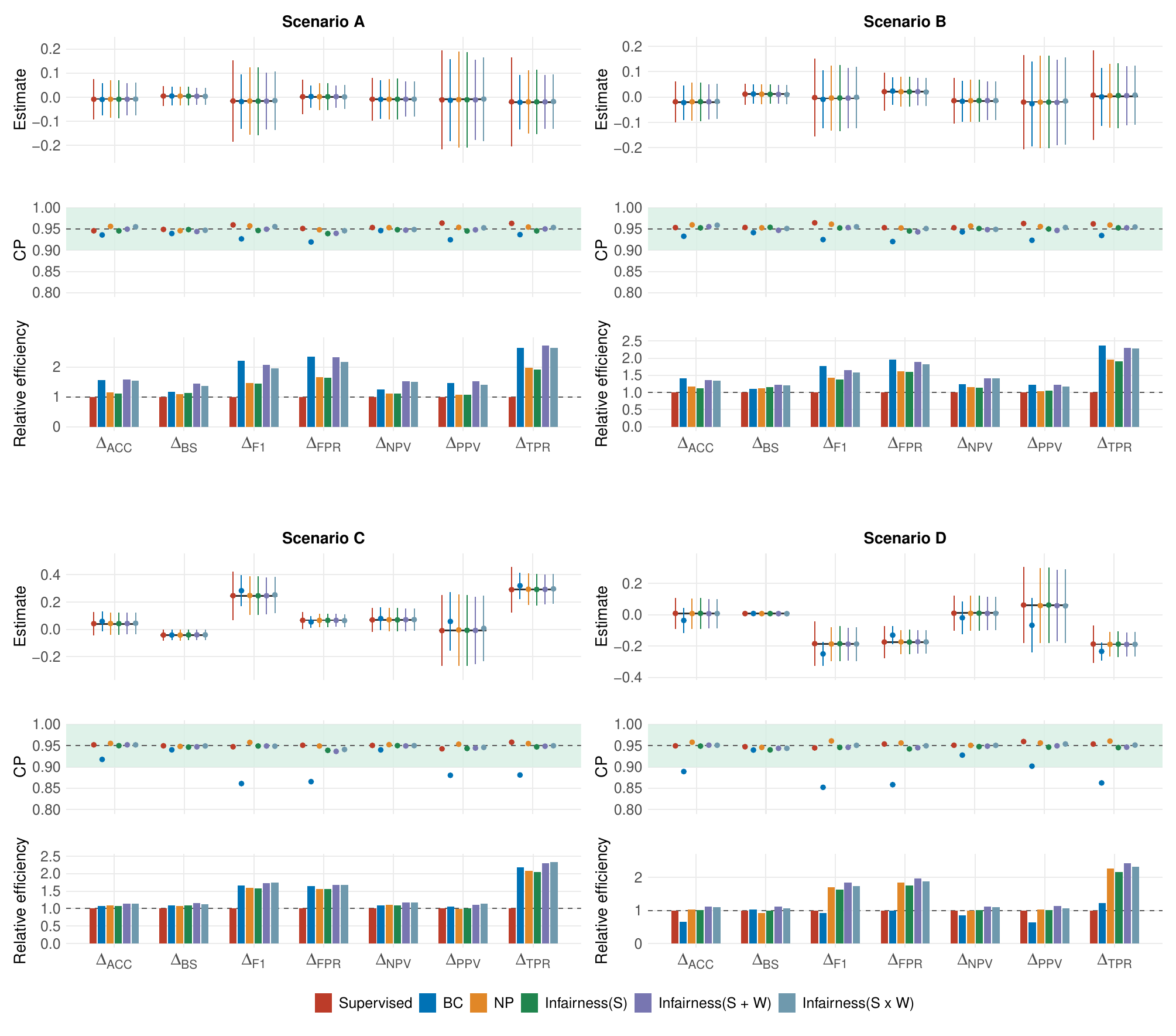}
\caption{\blue{\textbf{Group fairness auditing across simulation scenarios.} Results are based on $10^4$ Monte Carlo replicates with 400 labeled  and 20,000 unlabeled observations per replicate. Each panel corresponds to one simulation scenario. For each scenario, the three rows report point estimates, empirical coverage probability (CP), and relative efficiency (RE) for group differences in the performance metrics. In the point estimate row, vertical bars indicate approximate 95\% intervals based on empirical standard errors, and black horizontal bar mark the oracle values. In the CP row, the dashed line marks nominal 95\% coverage and the shaded region indicates coverage between 0.90 and 1.00. In the RE row, RE is defined as the ratio of the supervised mean squared error (MSE) to the MSE of each method. }} 
\label{fig:sim}
\end{figure}

\blue{The results are presented in Figure \ref{fig:sim} and are consistent with the take-aways of the stylized simulation study. Specifically, across all scenarios, the supervised estimator, NP, and the Infairness estimators exhibit minimal bias for all metrics and maintain coverage close to the nominal level. In Scenarios A and B where model misspecification is less severe, BC is also approximately unbiased, though with slightly lower coverage. In Scenarios C and D, BC exhibits substantial bias and under-coverage. Analogously to the stylized setting, increasing the labeled sample size does not remove this bias. With $n=1{,}000$, BC remains biased and the CP drops to approximately 80-85\% in Scenario C and 69-81\% in Scenario D (Figure \ref{fig:sim-1000}). We therefore focus our comparisons of efficiency on the NP and Infairness methods.}

\blue{Both NP and Infairness generally improve upon the supervised estimator in all settings. However, the Infairness approaches that adjust for $\bW$ often provides additional gains over NP, particularly for $\Delta_{\textrm{F1}}$, $\Delta_{\fpr}$, and $\Delta_{\tpr}$. In Scenarios A and B,  Infairness($S\times W$) and Infairness($S + W$) provide gains of 16-37\% over NP and Infairness($S$), while in Scenarios C and D, the gains are approximately 10\%. This pattern reflects the benefit of incorporating auxiliary covariates $\bW$ when they contain information about $Y$ beyond $S$. It is also important to note that Infairness($S$) is generally no worse than NP and Infairness($S\times W$) is no worse than using Infairness($S + W$). This suggests that richer imputation models are generally beneficial when the labeled sample size is sufficient to support the added model complexity.}

\section{Real data analysis}\label{s:real}
We next apply our method to audit fairness in \blue{an EHR-based phenotyping model for depression \citep{gehrmann_comparing_2018}. We focus on equal opportunity as lower rates of disease detection and diagnosis are often found among high-risk individuals within certain subgroups and, in many use cases, the benefit of diagnosis often outweighs potential harm. However, as the choice of fairness metric is ultimately use-case dependent, we also consider predictive equality for purposes of illustration. It is worth noting that metrics such as PPV, NPV, and $F_1$-score depend on the disease prevalence, which differs across groups in these data, making them less directly comparable \citep{gao_what_2024}. Throughout these analyses, we compare the three estimators from Section \ref{s:sim} that provide consistent estimation of the fairness metrics: supervised, NP, and Infairness. For Infairness, the group-specific imputation models were selected using BIC among the three candidate models in Section \ref{supp:tuning}. An additional audit of a chest X-ray classifier is provided in Section \ref{supp:chest}.}

\blue{We consider depression phenotyping from discharge summaries from the Medical Information Mart for Intensive Care III (MIMIC-III), a publicly available database of de-identified EHR data for over 40,000 patients admitted to critical care units of the Beth Israel Deaconess Medical Center between 2001 and 2012 \citep{johnson_mimic-iii_2016}. We restrict our analysis to each patient's first admission and exclude patients classified as ``frequent flyers'' who have more than three admissions within a calendar year \citep{gehrmann_comparing_2018}. The gold-standard depression label is available for $n = 872$ patients while $N = 28{,}766$ patients are unlabeled. The ML model was trained with a weakly supervised approach and the details are provided in Section \ref{supp:real-data-training}. Fairness was audited across race (White: 73\% vs non-White: 27\%). The auxiliary covariates, $\bW$, included age, sex, marital status, and insurance type.} 

\begin{figure}
    \centering
    \includegraphics[width=\linewidth]{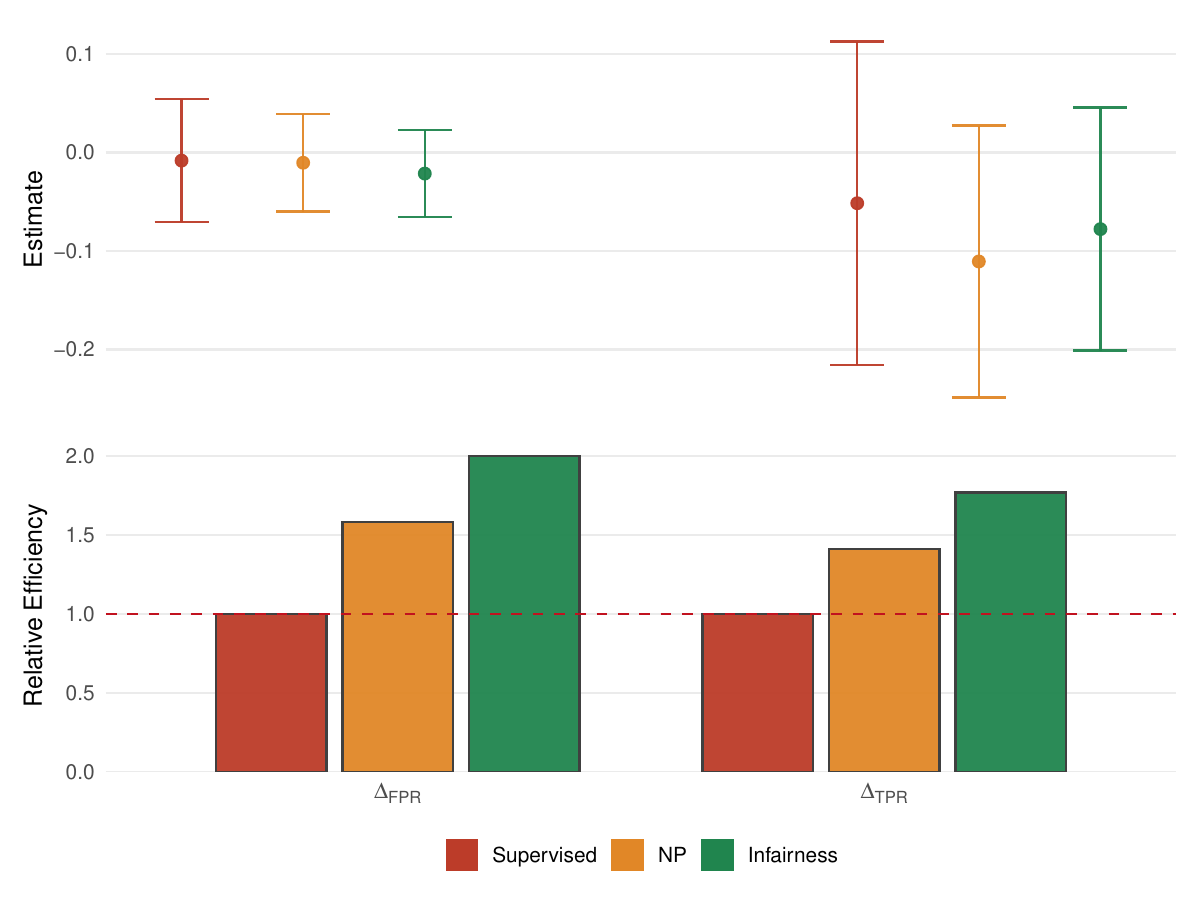}
    \caption{\blue{\textbf{FPR and TPR disparity estimates in the EHR phenotyping application across race.} Disparities are defined as the difference in group-specific performance between the non-White and White patients. }}
    \label{fig:mimic}
\end{figure}

\blue{Figure \ref{fig:mimic} shows results for equal opportunity and predictive equality. The point estimates from all three methods are generally within a reasonable range of one another and the 95\% confidence intervals suggest that there is no statistically significant difference in performance across groups. The main difference lies in precision: NP and Infairness substantially reduce variance relative to the supervised estimator, with Infairness providing the largest reduction. The RE of the Infairness estimators of $\Delta_{\tpr}$ and $\Delta_{\fpr}$ exceed 1.77, implying that our proposal reduces variance by at least 43.5\% relative to the supervised estimator. Practically, this translates into Infairness requiring 43.5\% fewer labeled observations than the supervised estimator to achieve the same precision. This is a 25\% improvement over the efficiency gains afforded by NP, which relies solely on $S$ for imputation.}

\section{Discussion}\label{s:diss}
Motivated by ML in biomedical applications, we proposed a SS method, called Infairness, to audit group fairness when labeled data is limited, but unlabeled data is abundant. Infairness uses a carefully constructed imputation model to enable evaluation of fairness criteria using both the labeled and unlabeled data. \blue{Our theoretical and empirical results highlight a simple message. With a sufficiently flexible imputation model, Infairness can yield more efficient inference than its supervised counterpart and often outperforms existing comparators when $\bW$ contains information about $Y$ beyond $S$. Importantly, we did not observe efficiency loss relative to existing approaches, suggesting the utility of Infairness for real-world auditing.} We close by commenting on several related issues and potential extensions. 

First, although our primary focus is on inference for difference-based disparity metrics as they are most commonly used in practice, Infairness is also applicable to to overall performance metrics as well as ratio-based metrics (e.g., $\frac{\Msc_0}{\Msc_1}$) using inference procedures derived from Theorem \ref{thm:ss-if}. In addition, Infairness supports hypothesis testing for $\epsilon$-fairness, a relaxed fairness criterion allowing group-specific disparities to fall within a small threshold, $\epsilon$ \citep{denis_fairness_2023}. Second, many protected attributes define more than two groups, and methods for evaluating multi-group fairness remain an active area of research. Early approaches relied on visualizations of model performance across groups, using tools such as Aequitas \citep{ saleiro_aequitas_2019}. Recent work has proposed meta-metrics, such as the max-min difference, max absolute difference, and variance of group-specific performance metrics. However, \citet{lum_-biasing_2022} showed that supervised estimators for these meta-metrics are biased. \citet{cherian_statistical_2024} later framed multi-group fairness as a multiple hypothesis testing problem, introducing a bootstrap-based method to jointly bound disparities across groups. Developing SS approaches for multi-group settings is a focus of our ongoing work.

Third, while our emphasis has been on group fairness, individual fairness and causal fairness are emerging within biomedical applications. Individual fairness requires that similar individuals receive similar predictions, while causal fairness seeks to uncover causal relationships between protected attributes and model outputs \citep[e.g.,][]{castiglione_faux_2022}. Extending our framework to support these fairness notions is an avenue of future research. Fourth, while we consider incorporating auxiliary covariates into the imputation step, including features used for training the ML model may further improve efficiency, particularly when the ML model is misspecified and the number of features is not large. However, care must be taken to avoid overfitting and a resulting loss in the efficiency gain of the SS estimator when the group-specific labeled sample sizes are not large. \blue{Additionally, small group-specific sample sizes can impact the variance estimates of Infairness. In Section \ref{supp:uninformative}, we conducted a simulation study with  $n = 100$ and the smallest group-specific sample size around 40.  While Infairness does have reduced coverage, its performance is no more impacted than supervised estimation or alternative SS approaches (Figure \ref{fig:small}).}

\blue{Finally, Infairness relies on a design-based choice that labeled data are obtained as a simple random sample from the underlying data pool. In practice, alternative sampling strategies can provide more efficient use of labeling resources. For example, labels can instead be sampled uniformly at random within levels of $A$ for fixed values of $n_0$ and $n_1$, in which case the proposed method remains valid. In other applications, the probability of observing labels may depend on the fully observed variables (e.g., $(\bX\trans, \bW\trans, A)\trans$), implying that $Y$ is missing at random. Approaches that adjust for selection bias, such as inverse probability weighting, would be required to obtain valid inference. Ongoing work in SS inference addresses more general missingness mechanisms and incorporating these advances into our current framework is a practically useful direction \citep[e.g.,][]{zhang2023double}}.


\backmatter


\section*{Acknowledgements}
J. Gronsbell is grateful for support of an NSERC Discovery Grant (RGPIN-2021-03734) and a Methodologist Seed Funding Grant from the University of Toronto Data Science Institute.


%


\putbib[references]



\section*{Supporting Information}
The supplementary material contains proofs, additional empirical results \blue{and code for reproducibility.}




\label{lastpage}

\end{bibunit}

\clearpage
\hypersetup{pageanchor=false}
\setcounter{page}{1}
\setcounter{footnote}{0}
\setcounter{section}{0}
\setcounter{subsection}{0}
\setcounter{subsubsection}{0}
\setcounter{figure}{0}
\setcounter{table}{0}
\setcounter{equation}{0}
\setcounter{theorem}{0}
\setcounter{lemma}{0}
\setcounter{corollary}{0}
\renewcommand{\thefigure}{S\arabic{figure}}
\renewcommand{\thetable}{S\arabic{table}}
\renewcommand{\thesection}{S\arabic{section}}
\renewcommand{\thetheorem}{S\arabic{theorem}}
\renewcommand{\theHsection}{supp.\arabic{section}}
\renewcommand{\theHsubsection}{supp.\arabic{section}.\arabic{subsection}}
\renewcommand{\theHsubsubsection}{supp.\arabic{section}.\arabic{subsection}.\arabic{subsubsection}}
\renewcommand{\theHfigure}{supp.\arabic{figure}}
\renewcommand{\theHtable}{supp.\arabic{table}}
\renewcommand{\theHequation}{supp.\arabic{equation}}
\renewcommand{\theHtheorem}{supp.\arabic{theorem}}
\renewcommand{\theHcorollary}{supp.\arabic{corollary}}
\renewcommand{\mhat}{\hat{m}}
\renewcommand{\bthetabar}{\bar{\btheta}}

\begin{bibunit}

\section{\blue{Implementation Details of SS Estimators}}

\subsection{\blue{Beta Calibration} Method of \citep{ji_can_2020} } \label{supp:beta-calibration}
We implement a frequentist version of the method proposed by \cite{ji_can_2020}. Specifically, for each group $ A = a $, we apply Beta calibration (BC) \citep{kull_beta_2017} to recalibrate the predictions $S$ by fitting the following logistic regression model on the labeled data:
$$
\text{logit}\{P(Y = 1 \mid S, A =a)\} = \zeta_{0,a} + \zeta_{1,a} \log S + \zeta_{2,a} \log(1 - S).
$$
Let $\hat{\zeta}_{0,a},\hat{\zeta}_{1,a},\hat{\zeta}_{2,a}$ denote the estimated coefficients. The calibrated prediction for group $A =a $ is
$
S^{\textrm{Beta}}_a = \text{expit}\{\hat{\zeta}_{0,a}+\hat{\zeta}_{1,a}\log S  + \hat{\zeta}_{2,a} \log (1-S)\} 
$. We then use $S^{\textrm{Beta}}_a$ to impute the missing outcome in group $A = a$. Again using TPR as an example, the corresponding estimator of $\Delta_{\tpr}$ is
\begin{align*}
   \widehat{\Delta}^{\text{\blue{BC}}}_{\tpr} & \coloneq \widehat{\tpr}^{\text{\blue{BC}}}_0 - \widehat{\tpr}^{\text{\blue{BC}}}_1 \\
   &= \frac{N_0^{-1}\sum_{i=n+1}^{N} D_iS^{\textrm{Beta}}_{0,i} I(A_i = 0)}{N_0^{-1}\sum_{i=n+1}^{N} S^{\textrm{Beta}}_{0,i} I(A_i = 0)}-\frac{N_1^{-1}\sum_{i=n+1}^{N} D_iS^{\textrm{Beta}}_{1,i} I(A_i = 1)}{N_1^{-1}\sum_{i=n+1}^{N} S^{\textrm{Beta}}_{1,i} I(A_i = 1)}.
\end{align*}

\subsection{\blue{\blue{Nonparametric Method of} \citet{gronsbell_semi-supervised_2018}}}\label{supp:kernel-smoothing}

\blue{We extend a group-specific version of the \blue{nonparametric (NP) approach of} \citet{gronsbell_semi-supervised_2018}. Specifically, for each group $A=a$, we estimate the conditional risk function $m_a(s) = \E(Y\mid S,A=a)$ with kernel smoothing as
$$
\widehat{m}_a(s)
=
\frac{
\sum_{i=1}^n K_{h_a}(S_i-s)Y_i I(A_i=a)
}{
\sum_{i=1}^n K_{h_a}(S_i-s) I(A_i=a)
},
$$
where $K_{h_a}(u)=h_a^{-1}K(u/h_a)$ is a given smooth, symmetric kernel function, $h_a= h(n_a)$ is a
bandwidth controlling the level of smoothing, and $nh_a^2 \to \infty$ and $nh_a^4 \to 0$ as $n_a\to \infty$. This \textit{under-smoothed} bandwidth is needed to ensure consistency and the authors recommend to use $h_a= \sigma_a /n_a^{-0.45}$, with $\sigma_a$ is the standard deviation of $S$ within group $A = a$.  }

\blue{We use $\widehat{m}_a(\cdot)$ to impute the missing outcomes in group $A=a$. Again using TPR as an example, the corresponding estimator of $\Delta_{\tpr}$ is
\begin{align*}
   \widehat{\Delta}^{\text{NP}}_{\tpr}
   &\coloneq
   \widehat{\tpr}^{\text{NP}}_0
   -
   \widehat{\tpr}^{\text{NP}}_1 \\
   &=
   \frac{
   N_0^{-1}\sum_{i=n+1}^{n+N} D_i \widehat{m}_0(S_i) I(A_i=0)
   }{
   N_0^{-1}\sum_{i=n+1}^{n+N} \widehat{m}_0(S_i) I(A_i=0)
   }
   -
   \frac{
   N_1^{-1}\sum_{i=n+1}^{n+N} D_i \widehat{m}_1(S_i) I(A_i=1)
   }{
   N_1^{-1}\sum_{i=n+1}^{n+N} \widehat{m}_1(S_i) I(A_i=1)
   }.
\end{align*}} 

\subsection{\blue{Basis Construction for Infairness}}\label{supp:tuning}
\blue{
We construct estimates of the adjusted and unadjusted conditional risk function using a natural cubic spline basis expansion of $S$ \citep{hastie_basis_2009}. Let $\boldsymbol{N}_K(S) = (N_1(S), \ldots, N_{K+2}(S))^\top$ denote a natural cubic spline basis of dimension $K+2$ constructed using a set of knots 
$\kappa_1 < \cdots < \kappa_K$. Such basis functions are piecewise cubic with continuous first and second derivatives and are constrained to be linear beyond the boundary knots.}
\blue{
We consider three candidate basis constructions for the group-specific imputation model:
\begin{align*}
\Bsc_{a}^{(1)}(S,\bW)
&=
\bigl(1, D, \boldsymbol{N}_K(S)^\top\bigr)^\top, \\
\Bsc_{a}^{(2)}(S,\bW)
&=
\bigl(1, D, \boldsymbol{N}_K(S)^\top, \bW^\top\bigr)^\top, \\
\Bsc_{a}^{(3)}(S,\bW)
&=
\bigl(1, D, \boldsymbol{N}_K(S)^\top, \bW^\top, (\boldsymbol{N}_K(S)\otimes \bW)^\top\bigr)^\top,
\end{align*}
where $\boldsymbol{N}_K(S)\otimes \bW$ denotes all pairwise products between the spline basis functions of $S$ and the components of $\bW$.}
\blue{
For $r \in \{1,2,3\}$, the corresponding imputation model is
\[
m_a^{(r)}(S,\bW)
=
g\!\left\{
(\bthetahat_a^{(r)})^\top \Bsc_a^{(r)}(S,\bW)
\right\},
\]
where $g(u) = \{1+\exp(-u)\}^{-1}$ is the logistic inverse link.}
\blue{
These three candidate models correspond to increasing levels of flexibility:
\begin{itemize}
\item Infairness($S$): nonlinear dependence on $S$ only
\item Infairness($S + \bW$): nonlinear effect of $S$ with additive effects of $\bW$
\item Infairness($S\times\bW$): additive effects together with interactions between $S$ and $\bW$.
\end{itemize}
}
\blue{
Throughout our simulation and real data, we chose $K = 3$ equally spaced knots. In our simulation study, we present results for the three versions of Infairness. In the real data, we select the model with the lowest BIC defined as:
$$-2\ell(\widehat\btheta_a)+d_a\,\min\{n_a^{0.1},\log(n_a)\},$$
where $\ell(\cdot)$ is the logistic log-likelihood, $d_a$ is the number of regression parameters, and $n_a$ is the group-specific labeled sample size.
}

\subsection{\blue{Variance Estimation for Infairness}}\label{supp:var-est}

\blue{To improve the finite-sample performance of the influence-function based variance estimation, we use $k$-fold cross validation. We detail the procedure for TPR below.}

\blue{For each group $a$, we randomly divide the labeled observations with $A=a$ into $k$ approximately equal folds. Let $\mathcal{I}_a^{k}$ denote the indices of the observations in the $k$th fold and $\mathcal{L}_a \setminus \mathcal{I}_a^{k}$ denote the indices of the observations not in the $k$th fold.  For fold $k$, we fit the imputation model using observations with indices in $\mathcal{L}_a \setminus \mathcal{I}_a^{k}$ to obtain $\widehat{\btheta}_a^{(-k)}$. We then calculate the corresponding Infairness estimator of the group-specific TPR as
$$ \left(\widehat{\tpr}^{\ssv}\right)^{(-k)}_a = \frac{N_a^{-1}\sum_{i=n+1}^{n+N}D_i\mhat^{(-k)}_{a,i} I(A_i = a)}{N_a^{-1}\sum_{i=n+1}^{n+N} \mhat^{(-k)}_{a,i}  I(A_i = a)}$$
where $\mhat^{(-k)}_{a,i}$ is the imputed value using $\widehat{\btheta}_a^{(-k)}$. The influence function contribution for the $i$th observation is then computed as  
$$
(\hat\mu_a^Y)^{-1}\{Y_i - \mhat^{(-k)}_{a,i}\}\left\{D_i - \left(\widehat{\tpr}^{\ssv}\right)^{(-k)}_a \right\} \mathbb{I}(i \in  \mathcal{I}_a^{k}).
$$
We apply the same procedure for all metrics as well as all the supervised, BC, and NP methods.}

\section{Asymptotic Analysis}
As described in Section \ref{s:pre} of the main text, we make no assumptions on the ML model, $\fhat$, other than that it is trained on an independent dataset, denoted as $\Dsc_{\train}$. For all the group-specific performance metrics, the target parameters are defined conditional on $\Dsc_{\train}$.  For example, the group-specific TPR is
\begin{equation*}
\tpr_a = \mathbb{P}\left(D = 1 \mid Y = 1, A = a\right)     
\end{equation*}  
where the probability is taken over the distribution of $(Y, \bX\trans, A)\trans$ conditioning on $\mathcal{D}_{\textrm{train}}$.  To simplify notation, we omit explicit conditioning on $\mathcal{D}_{\textrm{train}}$, with the understanding that all target parameters and corresponding inference procedures are defined conditionally on the training data.

\subsection{Regularity Conditions}\label{supp:ass}
To facilitate our discussion of asymptotic properties, we assume the following regularity conditions hold. 

\begin{condition}\label{cond:away}
Both $\mu_a^Y$ and $\mu_a^S$ are bounded away from 0 and 1.
\end{condition}

\begin{condition}\label{cond:compact}
(A) The basis, $\Bsc_a$, contains $S$, $D$, and an intercept, has compact support, and is of fixed dimension.  (B) The density function of  $\Bsc_a$ is continuously differentiable in its continuous components.  (C) There is at least one non-zero component of $\bthetabar_a$ and the link function is smooth. 
\end{condition}

\begin{condition}\label{cond:uniq}
There is no vector ${\btheta}_a$ such that $\mathbb{P}(\btheta_a \trans \Bsc_{a,i} >  \btheta_a \trans \Bsc_{a,j} \mid Y_i > Y_j ) = 1$ and $\mathbb{E} \left[\Bsc_{a} \Bsc_{a}\trans \dot{g}(\bthetabar_a\trans  \Bsc_{a}) \right] $ is positive definite.  
\end{condition}

Condition \ref{cond:away} ensures that the target parameters can be appropriately defined.  Condition \ref{cond:compact} is a standard assumption in Z-estimation theory while condition \ref{cond:uniq} ensures that there is no ${\btheta}_a$ that can perfectly separate the data based on $Y$ within each group defined by $A =a$ \citep{vaart_asymptotic_1998, tian_model_2007}.

\subsection{Asymptotic Properties of Supervised Estimators}\label{supp:sup}
Let $\muhat_a^Z = n_a^{-1}\sum_{i=1}^n Z_i I(A_i = a)$ denote the supervised estimator of $\mu_a^Z = \E[Z\mid A = a]$ for $Z \in \{Y, D, S^2, SY, DY\}$.  We first provide a simple Lemma that will be useful in establishing the asymptotic linear expansion of $\widehat{\Delta}_{\Msc}^{\supv}$.

\begin{lemma}\label{lemma:muz-sup}
Under the assumption that $n^{-1}\sum_{i=1}^n I(A_i = a) = n_a/n \overset{p}{\to} \rho_a \in (0, 1)$ as $n \to \infty$ from Section \ref{s:set-up},
$$
n^{\frac{1}{2}}\left(\muhat_a^Z - \mu_a^Z\right) = n^{-\frac{1}{2}}\rho_a^{-1}\sum_{i=1}^n \left\{(Z_i - \mu_a^Z) I(A_i = a) \right\} + o_p(1).
$$ 
\end{lemma}

\begin{proof}
The result is a direct consequence of the assumption that $n_a/n \overset{p}{\to} \rho_a$ as $n \to \infty$.  That is, 
\begin{align*}
n^{\frac{1}{2}}\left(\muhat_a^Z - \mu_a^Z\right)  &= n^{\frac{1}{2}}\left( \frac{ n^{-1} \sum_{i=1}^n (Z_i - \mu_a^Z) I(A_i = a)}{n^{-1} \sum_{i=1}^n  I(A_i = a)} \right) \\ 
&= n^{-\frac{1}{2}} \rho_a^{-1}  \sum_{i=1}^n (Z_i - \mu_a^Z)I(A_i = a) + o_p(1)
\end{align*}
where the second equality follows from Slutsky's Theorem.
\end{proof}

\subsubsection{Asymptotic Properties of $\Mschat^{\supv}_a$}
\begin{lemma}\label{lemma:M-sup}
Under the assumptions in Sections \ref{s:set-up} and \ref{supp:ass}, for any metrics $\Msc_a$ defined in Table \ref{tab:metrics}, $\Mschat^{\supv}_a \inp \Msc_a$ and 
$$
n^{\frac{1}{2}}\left(\Mschat^{\supv}_a - \Msc_a\right) =  n^{-\frac{1}{2}}\rho_a^{-1}\sum_{i=1}^n \textnormal{\bf IF}^{\supv}_{\Msc_a}(Y_i, S_i;c) I(A_i = a) + o_p(1)
$$
where $\textnormal{\bf IF}^{\supv}_{\Msc_a}(Y_i, S_i;c)$ is given in Table \ref{tab:if}. As $n\to\infty,  n^{\frac{1}{2}}\left(\Mschat^{\supv}_a - \Msc_a\right)$ converges weakly to a zero-mean Gaussian distribution with variance $$\rho_a^{-1} \E\left[ \left\{\textnormal{\bf IF}^{\supv}_{\Msc_a}(Y_i, S_i;c)\right\}^2\mid A = a\right].$$
\end{lemma}

\begin{proof}
The consistency of $\Mschat^{\supv}_a$ for $\Msc_a$ follows from the consistency of $\muhat^Z$ for $\mu^Z$ for $Z \in \{Y, D, S^2, SY, DY\}$.  We detail the derivation of the asymptotic expansion for $\tpr_a$. For all metrics $\Msc_a$ defined in Table \ref{tab:metrics}, the same techniques can be applied. By Lemma \ref{lemma:muz-sup}, 
$$
n^{\frac{1}{2}}(\muhat_a^{Y} - \mu_a^{Y}) = n^{-\frac{1}{2}}\rho_a^{-1}\sum_{i=1}^n (Y_i - \mu_a^{Y})I(A_i = a) + o_p(1)
$$
and
$$
n^{\frac{1}{2}}(\muhat_a^{DY} - \mu_a^{DY}) = n^{-\frac{1}{2}}\rho_a^{-1}\sum_{i=1}^n (D_iY_i - \mu_a^{DY})I(A_i = a) + o_p(1).
$$
Since $\mu_a^Y > 0$, a first-order Taylor expansion yields
\begin{align*}
    & \widehat{\tpr}^{\supv}_a = \frac{\muhat_a^{DY}}{\muhat_a^Y} = \frac{\mu_a^{DY}}{\mu_a^Y} + \rhoainv n^{-1}\sum_{i=1}^n 
    \begin{bmatrix}
    (\mu_a^Y)^{-1} & -(\mu_a^Y)^{-1}\tpr_a
    \end{bmatrix}
    \begin{bmatrix}
    (D_iY_i - \mu_a^{DY})I(A_i = a)\\
    (Y_i - \mu_a^{Y})I(A_i = a)
    \end{bmatrix} + o_p(n^{-\frac{1}{2}}).
\end{align*}
Therefore, 
\begin{align*}
    n^{\frac{1}{2}} \left(\widehat{\tpr}^{\supv}_a - \textrm{TPR}_a\right) & = n^{-\frac{1}{2}}\rho_a^{-1}\sum_{i=1}^n (\mu_a^Y)^{-1}\{Y_i (D_i - \tpr_a)\}I(A_i = a) + o_p(1)\\
    & = n^{-\frac{1}{2}}\rho_a^{-1}\sum_{i=1}^n \textnormal{\bf IF}^{\supv}_{\tpr_a}(Y_i, S_i;c) I(A_i = a) + o_p(1).
\end{align*}
By the standard Central Limit Theorem (CLT), $n^{\frac{1}{2}} \left(\widehat{\tpr}^{\supv}_a - \textrm{TPR}_a\right)$ converges weakly to a zero-mean Gaussian distribution with variance $\rho_a^{-1}\E\left[\left\{\textnormal{\bf IF}^{\supv}_{\tpr_a}(Y_i, S_i;c)\right\}^2 \mid A_i = a\right]$.
\end{proof}

\subsubsection{Asymptotic Properties of $\Deltahat^{\supv}_{\Msc}$}
We next apply Lemma \ref{lemma:M-sup} to find the asymptotic linear expansion of $\Delta_{\Msc}$.  

\begin{theorem}\label{thm:sup-if}
Under the assumptions in Sections \ref{s:set-up} and \ref{supp:ass}, $\Deltahat^{\supv}_{\Msc}\inp \Delta_{\Msc}$ and 
\begin{equation*}
 n^{\frac{1}{2}} \left(\Deltahat^{\supv}_{\Msc} - \Delta_{\Msc}\right)  =  \sum_{a \in \{0, 1\}} (-1)^{a}\rho_a^{-1}n^{-\frac{1}{2}}\sum_{i=1}^n \textnormal{\bf IF}^{\supv}_{\Msc_a}(Y_i, S_i;c) I(A_i = a) + o_p(1)
\end{equation*}
where $\textnormal{\bf IF}^{\supv}_{\Msc_a}(Y_i, S_i;c)$ is given in Table \ref{tab:if}. Therefore, $n^{\frac{1}{2}} \left(\Deltahat^{\supv}_{\Msc} - \Delta_{\Msc}\right)$ converges weakly to a zero-mean Gaussian distribution with variance $$\sum_{a\in\{0,1\}}\rho_a^{-1}\E\left[\left\{\textnormal{\bf IF}^{\supv}_{\Msc_a}(Y_i, S_i;c)\right\}^2\mid A = a\right].$$
\end{theorem}
\begin{proof}
By Lemma \ref{lemma:M-sup}, we have
$\Mschat^{\supv}_a \inp \Msc_a$ and 
$$
n^{\frac{1}{2}}\left(\Mschat^{\supv}_a - \Msc_a\right) = \rho_a^{-1} n^{-\frac{1}{2}}\sum_{i=1}^n \textnormal{\bf IF}^{\supv}_{\Msc_a}(Y_i, S_i;c) I(A_i = a) + o_p(1).
$$
Thus, $\Deltahat^{\supv}_{\Msc} \inp \Delta_{\Msc}$ and $\Mschat^{\supv}_a$ is also an asymptotically linear estimator with the following expansion
$$
 n^{\frac{1}{2}} \left(\Deltahat^{\supv}_{\Msc} - \Delta_{\Msc}\right)  =  \sum_{a \in \{0, 1\}} (-1)^{a}\rho_a^{-1}n^{-\frac{1}{2}}\sum_{i=1}^n \textnormal{\bf IF}^{\supv}_{\Msc_a}(Y_i, S_i;c) I(A_i = a) + o_p(1). 
$$
Therefore, the standard CLT implies $n^{\frac{1}{2}}\left(\Mschat^{\supv}_a - \Msc_a\right)$ converges weakly to a zero-mean Gaussian distribution with variance 
$
\sum_{a\in\{0, 1\}} \rho_a^{-1} \E\left[\left\{\textnormal{\bf IF}^{\supv}_{\Msc_a}(Y_i, S_i;c)\right\}^2 \mid A = a\right]. 
$
\end{proof}

\subsection{Asymptotic Properties of Semi-supervised Estimators}\label{supp:ss}

\subsubsection{Asymptotic Properties of \texorpdfstring{$\bthetahat_a$}{theta-hat a}}\label{supp:thetahata}
In the semi-supervised setting, we begin by considering the properties of $\bthetahat_a$, which solves the estimating equation  $\mathbb{Q}_a^{n_a}(\btheta_a;\lambda_{n_a}) \coloneq n_a^{-1}\sum_{i=1}^n \mathcal{B}_{a,i}\left\{Y_i - g\left(\btheta_{a}\trans \mathcal{B}_{a,i}\right)\right\}I(A_i = a) - \lambda_{n_a}\btheta_{a} = \bm{0},$ where $\Bsc_{a,i}=\Bsc_a(S_i,\bW_i)$.
\begin{lemma}\label{lemma:theta}
Let $\bthetabar_a$ be the solution to  $\mathbb{Q}_a(\btheta_a) = \E\left[ \Bsc_a\left\{Y - g\left(\btheta_{a}\trans \Bsc_a\right)\right\}I(A = a)\right]= \boldsymbol{0}.$ Under conditions \ref{cond:compact} -- \ref{cond:uniq}, $\bthetahat_a \overset{p}{\to} \bthetabar_a$. In addition, 
$$
n^{\frac{1}{2}}(\bthetahat_a - \bthetabar_a)  =  \rho^{-1}_a \bV_a^{-1} \left[n^{-\frac{1}{2}} \sum_{i=1}^n \Bsc_{a,i}\left\{Y_i - g\left(\bthetabar_a\trans \Bsc_{a,i}\right)\right\} I(A_i = a)\right] + o_p(1),
$$
where $\bV_a= \E\left\{\nabla g\left(\bthetabar_a\trans\Bsc_a\right)\Bsc_a\Bsc_a\trans\mid A =a \right\}$.  Therefore $n^{\frac{1}{2}}(\bthetahat_a - \bthetabar_a)$ converges weakly to a zero-mean Gaussian distribution with variance 
$$\rho_a^{-1}\bV_a^{-1} \E\left[\left\{Y-g\left(\bthetabar_a\trans\Bsc_a\right)\right\}^2 \Bsc_a\Bsc_a\trans \mid A=a\right] (\bV_a^{-1})\trans.$$
\end{lemma}

\begin{proof}
To show that $\bthetahat_a$  is consistent for $\bthetabar_a$ it suffices to verify that (i) $\sup_{\btheta_a}||\mathbb{Q}_a^{n_a}(\btheta_a;\lambda_{n_a})-\mathbb{Q}_a(\btheta_a)||_2 \overset{p}{\to} 0
$ and (ii) $\bthetabar_a$ is the unique solution to $\mathbb{Q}_a(\btheta_a) = \bm{0}$ \citep[Theorem 5.9,][]{vaart_asymptotic_1998}.  For (i), we note that $\lambda_{n_a} = o(n_a^{-\frac{1}{2}})$ and the class of functions, $$\left\{\Bsc_a\{Y - g\left(\btheta_a\trans \Bsc_a\right)\}I(A=a): \btheta_a \in \Theta_a \right\}$$ is uniformly bounded and hence a Glivenko-Cantelli class \citep[Theorem 19.4][]{vaart_asymptotic_1998}.  Under conditions \ref{cond:compact} -- \ref{cond:uniq} and again using the fact $\lambda_{n_a} = o(n_a^{-1/2})$, the arguments in Appendix A of \cite{tian_model_2007} imply that (ii) holds. It then follows that $\bthetahat_a \overset{p}{\to} \bthetabar_a$.

To establish the weak convergence of $n^{\frac{1}{2}}(\bthetahat_a - \bthetabar_a)$, note that a first-order Taylor expansion of $\mathbb{Q}_{a}^{n_a} (\bthetahat_a;\lambda_{n_a})$ about $\bthetabar_a$ together with the fact that $\lambda_{n_a} = o(n_a^{-\frac{1}{2}})$ implies that
$$
n^{\frac{1}{2}}(\bthetahat_a - \bthetabar_a)  =  \rho^{-1}_a \bV_a^{-1} \left[n^{-\frac{1}{2}} \sum_{i=1}^n \Bsc_{a,i}\left\{Y_i - g\left(\bthetabar_a\trans \Bsc_{a,i}\right)\right\} I(A_i = a)\right] + o_p(1). 
$$
The standard CLT implies $n^{\frac{1}{2}}(\bthetahat_a - \bthetabar_a)$ converges weakly to a zero-mean Gaussian distribution with variance 
$
\rho_a^{-1} \bV_a^{-1} \E\left[\left\{Y-g\left(\bthetabar_a\trans\Bsc_a\right)\right\}^2 \Bsc_a\Bsc_a\trans \mid A=a\right] (\bV_a^{-1})\trans.
$
\end{proof}

\subsubsection{Asymptotic Properties of $\mutilde_a^Z$}
We next derive the asymptotic expansions for the SS estimators of $\mu_a^Z$, denoted as $\mutilde_a^Z$, for $Z \in \{Y, D, SY, DY,S^2\}$.

\begin{lemma}\label{lemma:muz-ss}
Under the assumptions in Sections \ref{s:set-up} and \ref{supp:ass}, the following results hold:\\
\textbf{Case 1.} For \( Z \in \{D, S^2\} \), $
n^{\frac{1}{2}}\left(\widetilde{\mu}_a^Z - \mu_a^Z\right) \inp 0.$

\noindent \textbf{Case 2.} For $Z \in \{Y, SY, DY\} $, 
$$n^{\frac{1}{2}}\left(\mutilde_a^Z - \mu_a^Z\right) = \rhoainv \left\{n^{-\frac{1}{2}}\sum_{i=1}^n \left(Z_i - \Zbar_i\right)I(A_i = a)\right\} + o_p(1)$$
where 
$\Zbar_i = g\left(\bthetabar_a\trans\Bsc_{a,i}\right) \frac{Z_i}{Y_i} $.
Therefore, \( n^{\frac{1}{2}}\left(\widetilde{\mu}_a^Z - \mu_a^Z\right) \) converges in distribution to a zero-mean Gaussian distribution with variance: $
\rho_a^{-1} \E\left\{\left(Z - \Zbar\right)^2 \mid A = a\right\}.$
\end{lemma}
\begin{proof}
First, consider Case 1 in which the SS estimators of $\mu^Z$ do not require imputation.  Using the same arguments as the proof as in Lemma \ref{lemma:muz-sup}, it follows that 
$$N^{\frac{1}{2}}\left(\mutilde_a^Z - \mu_a^Z\right) = \rhoainv \left\{N^{-\frac{1}{2}}\sum_{i=1}^N (Z_i - \mu_a^Z)I(A_i = a)\right\} + o_p(1)$$
and hence $n^{\frac{1}{2}}\left(\mutilde_a^Z - \mu_a^Z\right) = o_p(1)$ 
given that $n /N \to 0$ as $n \to \infty$.

Next, consider Case 2 in which the SS estimators of $\mu^Z$ require imputation. We show the derivation for $Z = Y$, which can be applied for $Z \in \{SY, DY\}$.  We begin by rewriting   
\begin{align}
& n^{\frac{1}{2}}\left(\mutilde_a^Y - \mu_a^Y\right) \\
&= n^{\frac{1}{2}}\left[\left\{N_a ^{-1}\sum_{i=n+1}^{n+N} g\left(\bthetahat_a\trans\Bsc_{a,i}\right)I(A_i = a)\right\} - \mu_a^Y\right] \nonumber \\
&=  n^{\frac{1}{2}}\left[N_a^{-1}\sum_{i=n+1}^{n+N} \left\{g\left(\bthetahat_a\trans\Bsc_{a,i}\right) - g\left(\bthetabar_a\trans\Bsc_{a,i}\right)\right\}I(A_i = a)\right]  \nonumber \\
&\quad + \left(\frac{n}{N} \right)^{\frac{1}{2}} \left[ \rho_a^{-1} N^{-\frac{1}{2}} \sum_{i=n+1}^{n+N} \left\{g\left(\bthetabar_a\trans\Bsc_{a,i}\right) - \mu_a^Y\right\} I(A_i = a)\right] +o_p(1)  \nonumber\\ 
&= n^{\frac{1}{2}}\left[  \rho_a^{-1} N^{-1}\sum_{i=n+1}^{n+N} \left\{g\left(\bthetahat_a\trans\Bsc_{a,i}\right) - g\left(\bthetabar_a\trans\Bsc_{a,i}\right)\right\}I(A_i = a)\right] +o_p(1) \label{eq: mutilde_2}
\end{align}
The third equality follows from the assumption that $\lim_{n \to \infty} n/N = 0$ and the fact that 
$$\rho_a^{-1} N^{-\frac{1}{2}} \sum_{i=n+1}^{n+N} \left\{g\left(\bthetabar_a\trans\Bsc_{a,i}\right) - \mu_a^Y\right\} I(A_i = a) = O_p(1)$$
due to application of the standard CLT and using the fact that 
$$\E\left[ \left \{ Y - g\left(\bthetabar_a\trans\Bsc_a\right) \right\} I(A = a) \right] = 0$$
by definition of $\bthetabar_a$.  For the remaining term in (\ref{eq: mutilde_2}), the consistency of $\bthetahat_a$ for $\bthetabar_a$, together with a first-order Taylor expansion, implies
\begin{align}
& n^{\frac{1}{2}}\left[  \rho_a^{-1} N^{-1}\sum_{i=n+1}^{n+N} \left\{g\left(\bthetahat_a\trans\Bsc_{a,i}\right) - g\left(\bthetabar_a\trans\Bsc_{a,i}\right)\right\}I(A_i = a)\right] \nonumber \\
&=  \left\{\rho_a^{-1} N^{-1} \sum_{i=n+1}^{n+N} \nabla g\left(\bthetabar_a\trans\Bsc_{a,i}\right) \Bsc_{a,i}\trans I(A_i = a)\right\} \left\{n^{\frac{1}{2}}(\bthetahat_a - \bthetabar_a)\right\}  + o_p(1). \label{eq: mutilde-term2}
\end{align}
By Lemma \ref{lemma:theta}, 
$$
n^{\frac{1}{2}}(\bthetahat_a - \bthetabar_a)  =  \rho^{-1}_a \bV_a^{-1} \left[n^{-\frac{1}{2}}\sum_{i=1}^n \Bsc_{a,i}\left\{Y_i - g\left(\bthetabar_a\trans \Bsc_{a,i}\right)\right\} I(A_i = a)\right] + o_p(1).
$$
By the Weak Law of Large Numbers, 
\begin{align*}
\rho_a^{-1} N^{-1} \sum_{i=n+1}^{n+N} \left\{\nabla g\left(\bthetabar_a\trans\Bsc_{a,i}\right) \Bsc_{a,i}\trans I(A_i = a)\right\}  =  \E\left\{\nabla g\left(\bthetabar_a\trans\Bsc_a\right) \Bsc_a\trans \mid A = a\right\} + o_p(1). 
\end{align*}
To further simplify (\ref{eq: mutilde-term2}), we also note that
\begin{align*}
\E\left\{\nabla g\left(\bthetabar_a\trans\Bsc_a\right) \Bsc_a\trans \mid A = a\right\} \bV_a^{-1}  = \argmin_{\bbeta}\E\left\{\nabla g\left(\bthetabar_a\trans\Bsc_a\right)(1-\bbeta\trans\Bsc_a)^2\mid A = a\right\}  
\end{align*}
Since $\Bsc_a$ contains an intercept, the $\bbeta$ that minimizes the above expression is a vector with the first entry equal to 1 and the remaining entries equal to 0.  Therefore, $$\E\left\{\nabla g\left(\bthetabar_a\trans\Bsc_a\right) \Bsc_a\trans \mid A = a\right\} \bV_a^{-1} \Bsc_a = 1$$ and
\begin{align*}
n^{\frac{1}{2}}\left(\mutilde_a^Y - \mu_a^Y\right) &= \rhoainv\left[n^{-\frac{1}{2}}\sum_{i=1}^n \left\{Y_i - g\left(\bthetabar\trans_a\Bsc_{a,i}\right)\right\}I(A_i = a)\right] + o_p(1).
\end{align*}
It then follows by the standard CLT that $n^{\frac{1}{2}}\left(\mutilde_a^Y - \mu_a^Y\right)$ converges weakly to a zero-mean Gaussian distribution with variance $\rhoainv \E\left[\left\{Y - g\left(\bthetabar\trans_a\Bsc_a\right)\right\}^2\mid A = a\right]$.  
\end{proof}

\subsubsection{Asymptotic Properties of $\Mschat^{\ssv}_a$}
We use Lemma \ref{lemma:muz-ss} to derive the asymptotic properties of $\Mschat^{\ssv}_a$ in a similar manner to $\Mschat^{\supv}_a$.

\begin{lemma}\label{lemma:M-ss}
Under the assumptions in Sections \ref{s:set-up} and \ref{supp:ass},
$\Mschat^{\ssv}_a \inp \Msc_a$ and
$$
n^{\frac{1}{2}}\left(\Mschat^{\ssv}_a - \Msc_a\right) = \rho_a^{-1} n^{-\frac{1}{2}}\sum_{i=1}^n \textnormal{\bf IF}^{\ssv}_{\Msc_a}(Y_i, \Bsc_{a,i};c, \bthetabar_a) I(A_i = a) + o_p(1)
$$
where $\Bsc_{a,i}=\Bsc_a(S_i,\bW_i)$ and $\textnormal{\bf IF}^{\ssv}_{\Msc_a}(Y_i, \Bsc_{a,i};c, \bthetabar_a)$ is given in Table \ref{tab:if}. Therefore, $n^{\frac{1}{2}}\left(\Mschat^{\ssv}_a - \Msc_a\right)$ converges weakly to a zero-mean Gaussian distribution with variance $$\rho_a^{-1}\E\left[\left\{\textnormal{\bf IF}^{\ssv}_{\Msc_a}\left(Y_i, \Bsc_{a,i};c, \bthetabar_a\right)\right\}^2 \mid A = a\right].$$ 
\end{lemma}
\begin{proof}
The consistency of $\Mschat^{\ssv}_a$ for $\Msc_a$ follows from the consistency of $\mutilde^Z$ for $\mu^Z$ for $Z \in \{Y, D, S^2, SY, DY\}$.  Similar to the supervised estimator, we detail the derivation for $\tpr_a$ and note that all other metrics can be derived with similar arguments. By Lemma \ref{lemma:muz-ss},
$$
n^{\frac{1}{2}}(\mutilde_a^{Y} - \mu_a^{Y}) = \rhoainv \left[n^{-\frac{1}{2}}\sum_{i=1}^n \left\{Y_i - g\left(\bthetabar_a\trans\Bsc_{a,i}\right)\right\}I(A_i = a)\right] + o_p(1)
$$
and
$$
n^{\frac{1}{2}}(\mutilde_a^{DY} - \mu_a^{DY}) = \rhoainv \left[n^{-\frac{1}{2}}\sum_{i=1}^n D_i\left\{Y_i - g\left(\bthetabar_a\trans\Bsc_{a,i}\right)\right\}I(A_i = a)\right] + o_p(1).
$$
Similar to the proof of Lemma \ref{lemma:M-sup}, a first-order Taylor expansion yields
\begin{align*}
    & \frac{\mutilde_a^{DY}}{\mutilde_a^Y} = \frac{\mu_a^{DY}}{\mu_a^Y} + \rhoainv n^{-1}\sum_{i=1}^n 
    \begin{bmatrix}
    (\mu_a^Y)^{-1} & -(\mu_a^Y)^{-1}\tpr_a
    \end{bmatrix}
    \begin{bmatrix}
     D_i\left\{Y_i - g\left(\bthetabar_a\trans\Bsc_{a,i}\right)\right\}I(A_i = a)\\
     \left\{Y_i - g\left(\bthetabar_a\trans\Bsc_{a,i}\right)\right\}I(A_i = a)
    \end{bmatrix} + o_p(n^{-\frac{1}{2}}).
\end{align*}
Therefore, 
\begin{align*}
    n^{\frac{1}{2}} \left(\widehat{\tpr}_a^{\ssv} - \textrm{TPR}_a\right) & = \rho_a^{-1} \left[n^{-\frac{1}{2}}\sum_{i=1}^n (\mu_a^ Y)^{-1}\left\{Y_i - g\left(\bthetabar_a\trans\Bsc_{a,i}\right)\right\} (D_i - \tpr_a)I(A_i = a)\right] + o_p(1)\\
    & = \rho_a^{-1}\left\{n^{-\frac{1}{2}}\sum_{i=1}^n \textnormal{\bf IF}^{\ssv}_{\tpr_a}\left(Y_i, \Bsc_{a,i};c, \bthetabar_a\right) I(A_i = a)\right\} + o_p(1).
\end{align*}
By the standard CLT, $n^{\frac{1}{2}} \left(\widehat{\tpr}_a^{\ssv} - \textrm{TPR}_a\right)$ converges weakly to a zero-mean Gaussian distribution with variance $\rho_a^{-1}\E\left[\left\{\textnormal{\bf IF}^{\ssv}_{\tpr_a}\left(Y_i, \Bsc_{a,i};c, \bthetabar_a\right)\right\}^2 \mid A_i = a\right]$.
\end{proof}

\subsubsection{Asymptotic Properties of $\Deltahat^{\ssv}_{\Msc}$}
\renewcommand{\thetheorem}{\arabic{theorem}}
\renewcommand{\theHtheorem}{supp.main.\arabic{theorem}}
\setcounter{theorem}{0}
We conclude by providing derivations for the results in the main text.
\begin{theorem}[\textbf{Influence function of $\Deltahat^{\ssv}_{\Msc}$}]
Under the assumptions in Section \ref{s:set-up} and regularity conditions in Section \ref{supp:ass}, $\Deltahat^{\ssv}_{\Msc}\inp\Delta_{\Msc}$ and
\begin{equation*}
 n^{\frac{1}{2}} \left(\Deltahat^{\ssv}_{\Msc} - \Delta_{\Msc}\right)  =  n^{-\frac{1}{2}} \sum_{a \in \{0, 1\}} (-1)^{a}\rho_a^{-1}\sum_{i=1}^n \textnormal{\bf IF}^{\ssv}_{\Msc_a}\left(Y_i, \Bsc_{a,i};c, \bthetabar_a\right)I(A_i=a) +o_p(1)
\end{equation*}
where $\textnormal{\bf IF}^{\ssv}_{\Msc_a}(Y_i, \Bsc_{a,i};c, \bthetabar_a)$ is specified in Table \ref{tab:if}.  Therefore, $n^{\frac{1}{2}}\left(\Deltahat^{\ssv}_{\Msc} - \Delta_{\Msc}\right)$ converges weakly to a zero-mean Gaussian distribution with variance $$\sum_{a\in\{0,1\}}\rho_a^{-1}\E\left[\left\{\textnormal{\bf IF}^{\ssv}_{\Msc_a}\left(Y_i, \Bsc_{a,i};c, \bthetabar_a\right)\right\}^2\mid A = a\right].$$
\end{theorem}

\begin{proof}
The consistency of $\Deltahat^{\ssv}_{\Msc}$ for $\Delta_{\Msc}$ follows from Lemma \ref{lemma:M-ss}.
The asymptotic expansion for $\Deltahat^{\ssv}_{\Msc}$ similarly follows from Lemma \ref{lemma:M-ss} and is given by,
\begin{equation*}
 n^{\frac{1}{2}} \left(\Deltahat^{\ssv}_{\Msc} - \Delta_{\Msc}\right)  =  \sum_{a \in \{0, 1\}} (-1)^{a}\rho_a^{-1}\left\{n^{-\frac{1}{2}}\sum_{i=1}^n \textnormal{\bf IF}^{\ssv}_{\Msc_a}\left(Y_i, \Bsc_{a,i};c, \bthetabar_a\right)I(A_i=a) \right\}+o_p(1).
\end{equation*}
Using same argument as in the proof of Theorem \ref{thm:sup-if}, 
$n^{\frac{1}{2}}\left(\Deltahat^{\ssv}_{\Msc} - \Delta_{\Msc}\right)$ converges weakly to a zero-mean Gaussian distribution with variance 
$$\sum_{a\in\{0,1\}}\rho_a^{-1}\E\left[\left\{\textnormal{\bf IF}^{\ssv}_{\Msc_a}\left(Y_i, \Bsc_{a,i};c, \bthetabar_a\right)\right\}^2\mid A = a\right].$$
\end{proof}

\subsection{Asymptotic Variance Analysis of $\Deltahat^{\supv}_{\Msc}$ and $\Deltahat^{\ssv}_{\Msc}$} \label{supp:var-comp}

Theorems \ref{thm:sup-if} and \ref{thm:ss-if} establish that both estimators $\Deltahat^{\supv}_{\Msc}$ and $\Deltahat^{\ssv}_{\Msc}$ are asymptotically linear. However, in the influence function $\textnormal{\bf IF}^{\ssv}_{\Msc_a}\left(Y_i, \Bsc_{a,i}; c, \bthetabar_a\right)$ provided in Table \ref{tab:if}, $Y_i$ is centered at its imputed value $g(\bthetabar_a\trans\Bsc_{a,i})$. This difference in centering is responsible for the lower asymptotic variance of $\Deltahat^{\ssv}_{\Msc}$ relative to $\Deltahat^{\supv}_{\Msc}$.  We establish this result more formally in the following corollary. 

\begin{corollary}[\textbf{Variance comparison of $\Deltahat^{\supv}_{\Msc}$ and $\Deltahat^{\ssv}_{\Msc}$}]
When the imputation model in equation \eqref{eq:imputation} in the main text is correctly specified and $\E(Y\mid S, \bW, A = a) \ne \E(Y \mid A = a)$, 
$$
\sum_{a\in\{0,1\}}\rho_a^{-1}\E\left[\left\{\textnormal{\bf IF}^{\ssv}_{\Msc_a}(Y_i, \Bsc_{a,i};c, \bthetabar_a)\right\}^2\mid A = a\right] < \sum_{a\in\{0,1\}}\rho_a^{-1}\E\left[\left\{\textnormal{\bf IF}^{\supv}_{\Msc_a}(Y_i, S_i;c)\right\}^2\mid A = a\right].
$$
\end{corollary}

\begin{proof}
We again verify the result for $\Delta_{\tpr}$ and note that other metrics can be derived similarly. When the imputation model \eqref{eq:imputation} is correct, 
$$\textnormal{\bf IF}^{\ssv}_{\tpr_a}\left(Y_i, \Bsc_{a,i};c, \bthetabar_a\right) = (\mu_a^Y)^{-1}\{Y_i- \E(Y_i\mid S_i,\bW_i, A_i = a)\}\left(D_i - \tpr_a\right)$$
and $$\textnormal{\bf IF}^{\supv}_{\tpr_a}\left(Y_i, S_i;c\right) = (\mu_a^Y)^{-1}\{Y_i(D_i - \tpr_a)\}.$$ Thus,
\begin{align*}
&\E\left[\left\{\textnormal{\bf IF}^{\ssv}_{\tpr_a}\left(Y_i, \Bsc_{a,i};c, 
 \bthetabar_a\right)\right\}^2\mid A = a\right] \\
= & \E\left[(\mu_a^Y)^{-2}\{Y_i- \E(Y_i\mid S_i,\bW_i, A_i = a)\}^2\left(D_i - \tpr_a\right)^2\mid A = a\right]\\
= &\E\left[\left\{\textnormal{\bf IF}^{\supv}_{\tpr_a}\left(Y_i, S_i;c\right)\right\}^2\mid A = a\right] - \E\left[(\mu_a^Y)^{-2}\{E(Y_i\mid S_i,\bW_i, A_i = a)\}^2\left(D_i - \tpr_a\right)^2\mid A = a\right] \\
< & \E\left[\left\{\textnormal{\bf IF}^{\supv}_{\tpr_a}\left(Y_i, S_i;c\right)\right\}^2\mid A = a\right].
\end{align*}
The last inequality is strict due to regularity condition \ref{cond:away}. It follows that $$\sum_{a\in\{0,1\}}\rho_a^{-1}\E\left[\left\{\textnormal{\bf IF}^{\ssv}_{\tpr_a}\left(Y_i, \Bsc_{a,i};c, \bthetabar_a\right)\right\}^2\mid A = a\right] < \sum_{a\in\{0,1\}}\rho_a^{-1}\E\left[\left\{\textnormal{\bf IF}^{\supv}_{\tpr_a}\left(Y_i, S_i;c\right)\right\}^2\mid A = a\right].$$  We note the assumption that $\E(Y\mid S, \bW, A = a) \ne \E(Y \mid A = a)$ is only needed to establish this result for metrics based on the NPV and PPV. 
\end{proof}

\section{\blue{Simulation Studies}}

\subsection{\blue{Stylized Setting Data Generation}}\label{supp:sim-dgp-stylized}

\blue{
For the stylized setting, we generate a binary protected attribute
$$A \sim \mathrm{Bernoulli}(0.6),$$
and a binary auxiliary covariate
$$W \sim \mathrm{Bernoulli}(0.5).$$
The outcome is generated independently as
$$Y \sim \mathrm{Bernoulli}(0.3).$$
Conditional on $(Y,A,W)$, we generate the prediction  directly from a beta distribution as
$$S\mid Y=y,A=a,W=w \sim \mathrm{Beta}\{\alpha_{a w y},\beta_{a w y}\}.$$
The final binary prediction of $Y$ is $D=I(S\geq 0.5)$. We set $N = 10,0000$ and $n = 400$. In the first scenario, $S \perp W \mid Y = y, A = a$. In the second scenario, $S \not\perp W \mid Y = y, A = a$.}

\paragraph{\blue{Scenario 1}}
\blue{The parameters of the group-specific beta distributions are summarized in \ref{eq:sty-sc1}. 
\begin{align}
\begin{array}{c c c c c}
A & W & Y & \alpha_{a w y} & \beta_{a w y}\\
\hline
0,1 & 0,1 & 1 & 7 & 5\\
0,1 & 0,1 & 0 & 5 & 7\\
\end{array}\label{eq:sty-sc1}
\end{align}
Under this data-generating mechanism, the assumptions of the beta calibration (BC) method are satisfied as
\begin{align*}
\mathrm{logit}\{\E(Y\mid S,A, \bW)\} = \log(3/7)
   + 2\log S - 2\log(1-S).
\end{align*}
}
\paragraph{\blue{Scenario 2}}
\blue{
The distribution of $S$ within group $A=1$ remains the same as before, while the the distribution of $S$ within group $A=0$ depends on $W$. The parameters of the group-specific beta distributions are summarized in \ref{eq:sty-sc2}.
\begin{align}
\begin{array}{c c c c c}
A & W & Y & \alpha_{a w y} & \beta_{a w y}\\
\hline
0 & 0 & 1 & 14.0 & 4.0\\
0 & 1 & 1 & 4.5 & 10.0\\
0 & 0 & 0 & 2.5 & 10.0\\
0 & 1 & 0 & 7.0 & 5.0\\
1 & 0,1 & 1 & 7 & 5\\
1 & 0,1 & 0 & 5 & 7
\end{array}\label{eq:sty-sc2}
\end{align}
For group $A=0$, marginalizing over $W$ induces a relationship between $Y$ and $S$ that is not captured by standard BC:
\begin{align*}
\mathrm{logit}\{\E(Y\mid S,W,A=0)\}
&=\log(3/7)+11.5\log S-6\log(1-S)\\
&\quad-2.6W-14W\log S+11W\log(1-S).
\end{align*}}

\subsection{\blue{Traditional Setting Data Generation}}\label{supp:sim-dgp-traidtional}
\blue{The outcome $Y$ was generated under four different scenarios to vary the quality of the fit of the ML model. In scenarios A-C, the intercepts $\alpha_0$ and $\alpha_1$ are chosen such $P(Y = 1) \approx 0.3$. }

\subsubsection{\blue{Scenario A: Logistic model with nonlinear and interaction terms}}

\blue{The outcome $Y$ is generated according to a logistic model with group-specific coefficients such that 
\[
\begin{aligned}
P(Y=1 \mid \bX, \bW, A=0)
= \operatorname{expit}\big(
&\alpha_0
+ 1.9X_1 + 1.9X_2 + 0.9X_3 + 0.9X_4
+ 1.6(W_1+W_2+W_3) \\
&+ 0.4X_2^2 - 0.5X_3^3
+ 0.6X_5X_6 + 0.1X_2X_5X_6
\big), 
\\[0.5em]
P(Y=1 \mid \bX, \bW, A=1)
= \operatorname{expit}\big(
&\alpha_1
+ 1.7X_1 + 1.7X_2 + 0.7X_3 + 0.7X_4
+ 1.2(W_1+W_2+W_3) \\
&+ 0.4X_2^2 - 0.5X_3^3
+ 0.8X_5X_6 - 0.1X_3X_5X_6
\big),
\end{aligned}
\]
and $\operatorname{expit}(x) = (1 + e^{-x})^{-1}$. }

\subsubsection{\blue{Scenario B: Non-logistic activation function with non-linear hidden layer}}

\blue{
The outcome $Y$ is generated using a non-logistic activation function applied to the group-specific non-linear predictors such that
\[
\begin{aligned}
P(Y=1 \mid \bX, \bW, A=0)
= \exp\Big\{
-&\big(
\alpha_0
+ 0.40X_1 - 0.30X_2 + 0.15X_3 - 0.15X_4 \\
&+ 0.25W_1 - 0.20W_2 + 0.20W_3
\big)^2
\Big\}, \mbox{ and }
\\[0.5em]
P(Y=1 \mid \bX, \bW, A=1)
= \exp\Big\{
-&\big(
\alpha_1
+ 0.35X_1 - 0.25X_2 + 0.20X_3 - 0.20X_4 \\
&+ 0.15W_1 - 0.15W_2 + 0.20W_3
\big)^2
\Big\}.
\end{aligned}
\]
}

\subsubsection{\blue{Scenario C: Complementary log-log }}\label{supp:clog}
\blue{The outcome $Y$ is generated using a complementary log-log model with group-specific nonlinear transformations applied to the linear predictor that introduce local non-monotonicity into the resulting risk score. We first generate the overall linear predictor as  
\[
\begin{aligned}
\eta
={}& -0.5
+X_1+0.8X_2+0.6X_3+0.5X_4+0.3X_5 \\
&-0.1X_5X_2
+0.1X_3^2
-0.03X_3^3 \\
&+0.9W_1+0.6W_2-0.6W_3 \\
&+0.3X_1W_1-0.3X_2W_2+0.2X_3W_3
+0.2X_4W_1-0.2X_5W_2.
\end{aligned}
\]
We then standardize $\eta$ as $z  = \frac{\eta - \E(\eta)}{\text{sd}(\eta)}$ and obtain the group-specific event probabilities as
\[
\begin{aligned}
P(Y=1 \mid \bX,\bW,A=0)
=
1-\exp\Big[
-\exp\Big\{
&\alpha_0
+3.2z
+7.5\exp\Big(-\frac{(z-0.55)^2}{2(0.18)^2}\Big)
\Big\}
\Big], \mbox{ and }
\\[0.5em]
P(Y=1 \mid \bX,\bW,A=1)
=
1-\exp\Big[
-\exp\Big\{
&\alpha_1
+3.2z
+7.5\exp\Big(-\frac{(z+0.30)^2}{2(0.18)^2}\Big)
\Big\}
\Big].
\end{aligned}
\]
}

\subsubsection{\blue{Scenario D: Tree Structure}}\label{supp:tree}
\blue{
In the tree-based setting, the event probability is a piecewise-constant function of two linear predictors based on $\bX$ and $\bW$. Define
\[
\eta_{\bX} =
X_1 + X_2 - X_3
+ 0.35X_4 - 0.25X_5 + 0.2X_6
+ 0.15X_1X_2 - 0.1X_3X_4,
\]
and
\[
\eta_{\bW} =
0.85W_3 + 0.75W_4 - 0.75W_5
+ 0.15W_3W_4.
\]
The risk score $P( Y = 1 \mid \bX, \bW, A)$ is specified as a piecewise-constant function of $\eta_{\bX}$ and $\eta_{\bW}$, with values given in Table \ref{tab:tree}. The likelihood of the outcome increases across the $\eta_{\bW}$ intervals within each $\eta_{\bX}$ interval, but varies non-monotonically across $\eta_{\bX}$ intervals and differently within each protected group.}
\begin{table}[h]
\centering
\caption{\blue{$P( Y = 1 \mid \bX, \bW, A)$ in the tree-based setting.}}
\label{tab:tree}
\resizebox{\textwidth}{!}{%
\begin{tabular}{lccc|ccc}
\hline
& \multicolumn{3}{c|}{$A=0$} & \multicolumn{3}{c}{$A=1$} \\
\cline{2-7}
$\eta_{\bX}$ interval
& $\eta_{\bW} \le -0.8$
& $-0.8<\eta_{\bW}\le 1.1$
& $\eta_{\bW}>1.1$
& $\eta_{\bW} \le -1.2$
& $-1.2<\eta_{\bW}\le 0.8$
& $\eta_{\bW}>0.8$ \\
\hline
$\eta_{\bX} \le -3.0$        
& 0.50 & 0.74 & 0.94
& 0.03 & 0.23 & 0.47 \\

$-3.0 < \eta_{\bX} \le -1.7$ 
& 0.03 & 0.20 & 0.44
& 0.40 & 0.64 & 0.88 \\

$-1.7 < \eta_{\bX} \le -0.7$ 
& 0.46 & 0.70 & 0.94
& 0.03 & 0.20 & 0.44 \\

$-0.7 < \eta_{\bX} \le 0.4$  
& 0.03 & 0.24 & 0.48
& 0.42 & 0.66 & 0.90 \\

$0.4 < \eta_{\bX} \le 1.5$   
& 0.10 & 0.34 & 0.58
& 0.50 & 0.74 & 0.94 \\

$1.5 < \eta_{\bX} \le 3.2$   
& 0.46 & 0.70 & 0.94
& 0.03 & 0.23 & 0.47 \\

$\eta_{\bX} > 3.2$           
& 0.18 & 0.42 & 0.66
& 0.54 & 0.78 & 0.94 \\
\hline
\end{tabular}%
}
\end{table}

\clearpage

\subsection{\blue{Stylized Setting with $n= 1000$}}

\begin{figure}[htbp!]
    \centering
    \includegraphics[width=\linewidth]{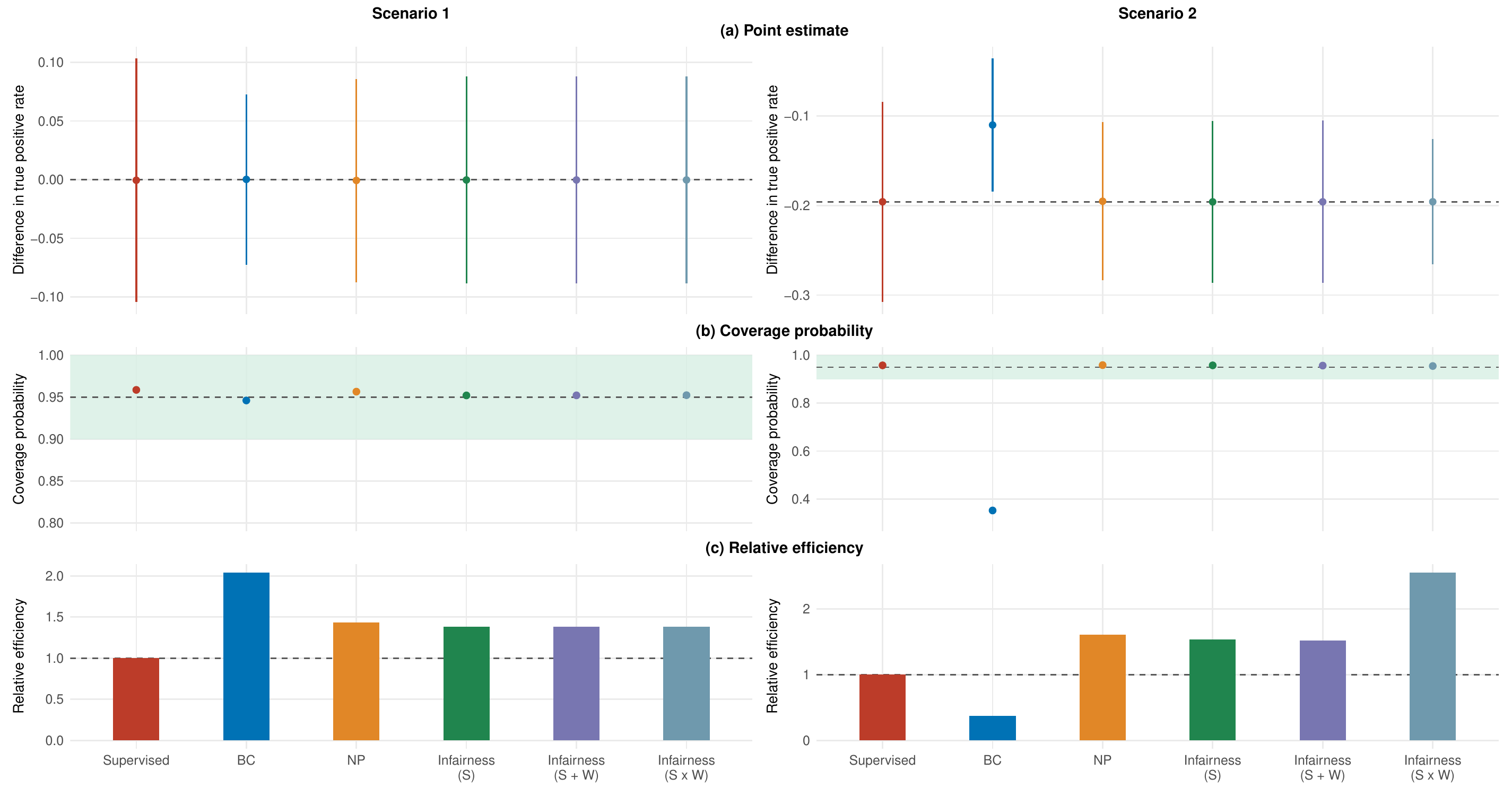}
    \caption{\blue{\textbf{Stylized setting for estimating the difference in true positive rates.} Results are based on $10^4$ Monte Carlo replicates with 20,000 validation observations and 1000 labeled observations per replicate. Panel (a) shows the average point estimate of \(\Delta_{\tpr}=\tpr_0-\tpr_1\), with vertical lines corresponding to the average estimate plus or minus 1.96 times the empirical standard error; the dashed horizontal line marks the truth. Panel (b) shows empirical coverage probability of nominal 95\% confidence intervals, with the shaded region indicating coverage between 0.90 and 1.00. Panel (c) shows relative efficiency, defined as the ratio of the supervised mean squared error to the mean squared error of each method. }} \label{fig:toy-1000}
\end{figure}

\clearpage

\subsection{\blue{Traditional Setting with $n = 1000$}}\label{supp:sim-large}

\begin{figure}[h!]
    \centering
    \includegraphics[width=\linewidth]{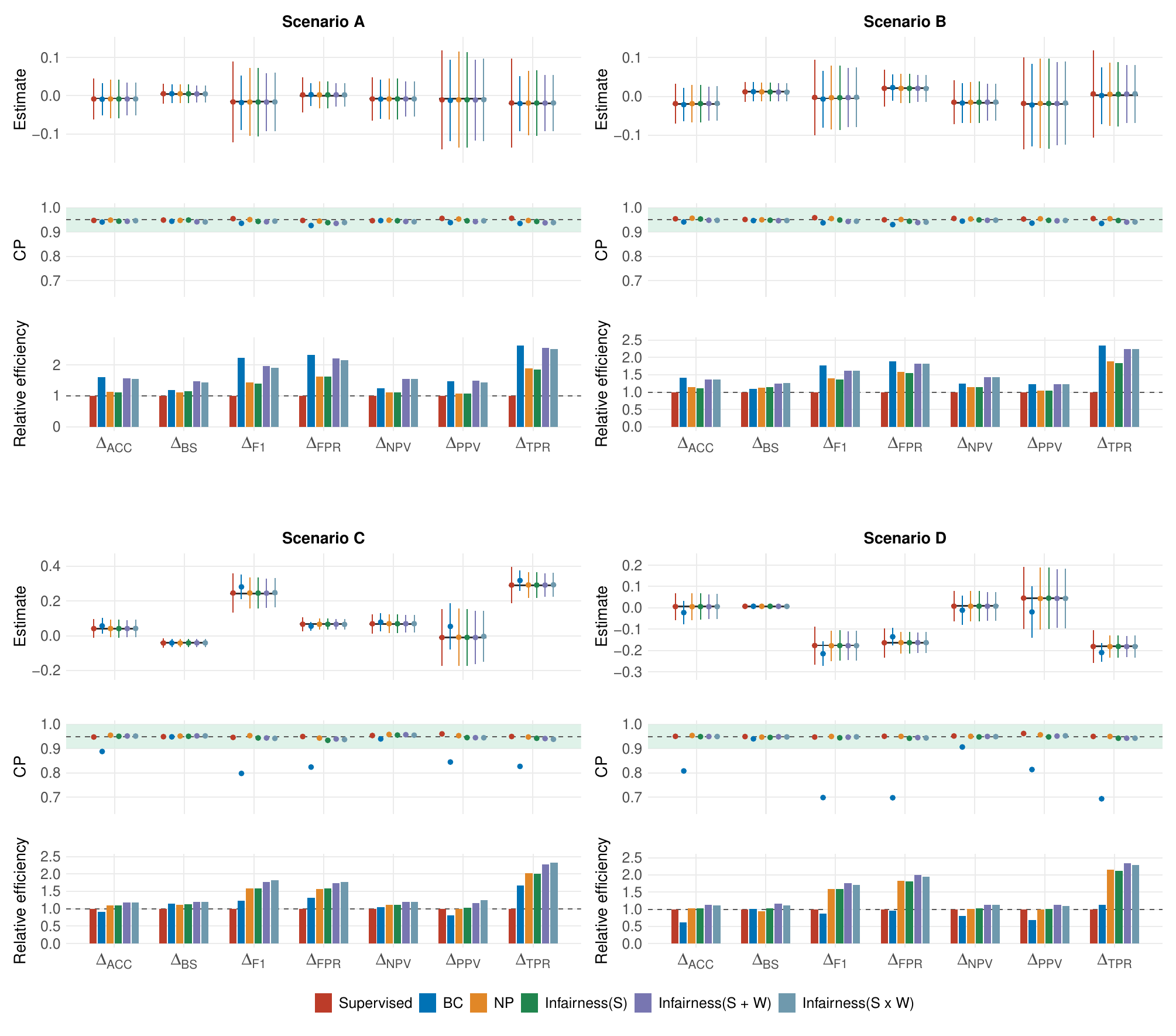}
\caption{\blue{\textbf{Group fairness auditing across simulation scenarios.} Results are based on $10^4$ Monte Carlo replicates with 1,000 labeled  and 20,000 unlabeled observations per replicate. Each panel corresponds to one simulation scenario. For each scenario, the three rows report point estimates, empirical coverage probability (CP), and relative efficiency (RE). In the point estimate row, vertical bars indicate approximate 95\% intervals based on empirical standard errors and the black horizontal bar is the true values. In the CP row, the dashed line marks nominal 95\% coverage and the shaded region indicates coverage between 0.90 and 1.00.}} 
\label{fig:sim-1000}
\end{figure}

\subsection{\blue{Small Sample Size $n = 100$}}
\begin{figure}[H]
    \centering
    \includegraphics[width=\linewidth]{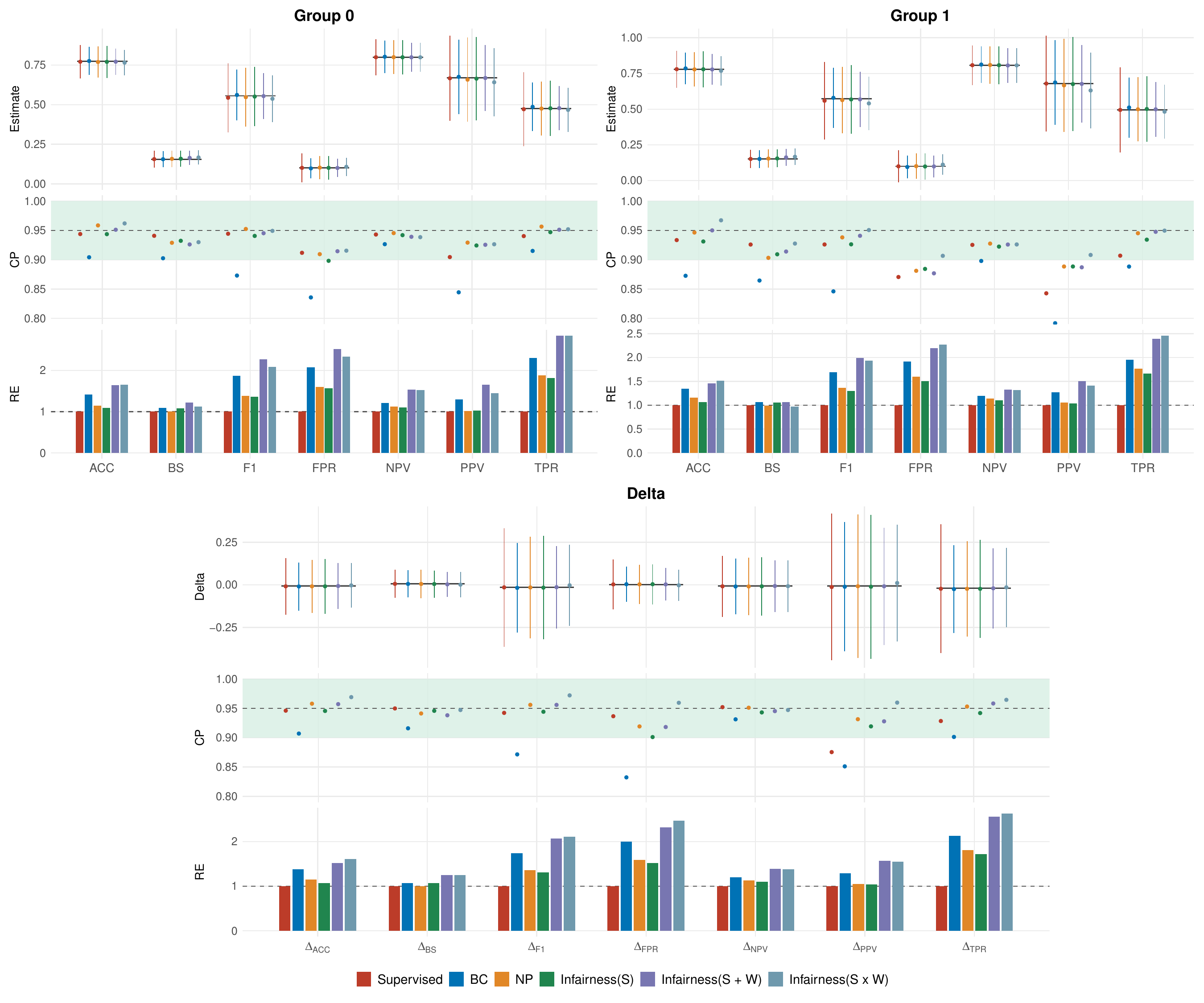}
\caption{\blue{\textbf{Group fairness auditing in Scenario A with $n = 100$.} Results are based on $10^4$ Monte Carlo replicates with 100 labeled and 20,000 unlabeled observations per replicate. Each panel corresponds to group-specific results or their differences. For each panel, the three rows report point estimates, empirical coverage probability (CP), and relative efficiency (RE) for group differences in the performance metrics. In the point estimate row, vertical bars indicate approximate 95\% intervals based on empirical standard errors and the black horizontal bar marks the truth. In the CP row, the dashed line marks nominal 95\% coverage and the shaded region indicates coverage between 0.90 and 1.00.}} 
\label{fig:small}
\end{figure}

\clearpage

\subsection{\blue{Uninformative prediction}}\label{supp:uninformative}
\begin{figure}[htbp!]
    \centering
    \includegraphics[width=\linewidth]{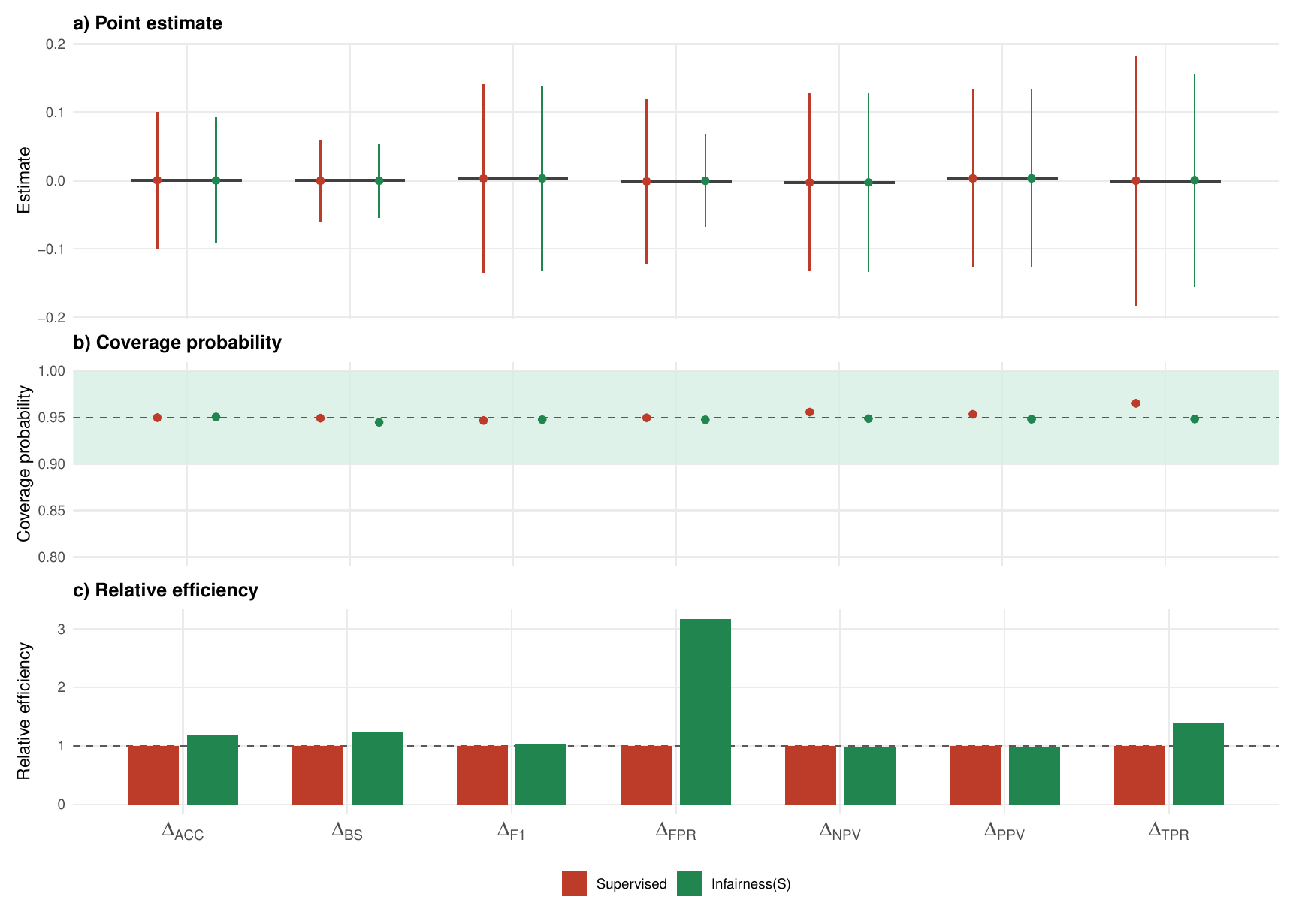}
\caption{\blue{\textbf{Group fairness auditing with predictions generated from Uniform[0,1].} Results are based on $10^4$ Monte Carlo replicates with 400 labeled  and 20,000 unlabeled observations per replicate. The three rows report point estimates, empirical coverage probability (CP), and relative efficiency (RE) for group differences in the performance metrics. In the point estimate row, vertical bars indicate approximate 95\% intervals based on empirical standard errors and the black horizontal bar marks the true values. In the CP row, the dashed line marks nominal 95\% coverage and the shaded region indicates coverage between 0.90 and 1.00. In the RE row, RE is defined as the ratio of the supervised mean squared error (MSE) to the MSE of each method.}} 
\label{fig:uninformative}
\end{figure}

\blue{As an extreme setting, we randomly generate the prediction from a Uniform[0,1] distribution. Results are presented in Figure \ref{fig:uninformative}. Our proposed method remains unbiased and has nominal coverage. With respective to efficiency, Infairness does not provide meaningful gains in $\Delta_{\ppv}, \Delta_{\npv}$ and $\Delta_{\fone}$, but does not lose efficiency either. Infairness can still gain in the remaining metrics, which is briefly detailed below for $\Delta_{\tpr}$ and $\Delta_{\ppv}$. }

\blue{When $S\indep Y$, our imputation model $\hat{m}_a \inp \mu_a^Y = \E(Y|A =a)$. The IF for the supervised TPR estimator is $(\mu_a^Y
)^{-1}\{Y_i (D_i - \tpr_a)\}$, while the IF for Infairness estimator is centered at $(\mu_a^Y
)^{-1}\{Y_i - \E(Y|A =a)\} (D_i - \tpr_a)$. This leads to variance reduction in estimating $\tpr_a$ as
\[
\begin{aligned}
\textrm{aVar}(\widehat{\tpr}^{\sup}_a) &=
(\mu_a^Y)^{-2}
\E\left[
Y_i^2(D_i-\tpr_a)^2
\mid A_i=a
\right] \\
&=
(\mu_a^Y)^{-2}
\E(Y_i\mid A_i=a)
\E\left[
(D_i-\tpr_a)^2
\mid A_i=a
\right] \\
& =
\frac{1}{\mu_a^Y}
\tpr_a(1-\tpr_a)\\
\end{aligned}
\]
while
\[
\begin{aligned}
\textrm{aVar}(\widehat{\tpr}^{\ssv}_a) &
=
(\mu_a^Y)^{-2}
\E\left[
(Y_i-\mu_a^Y)^2
\mid A_i=a
\right]
\E\left[
(D_i-\tpr_a)^2
\mid A_i=a
\right] \\
& =
\frac{1-\mu_a^Y}{\mu_a^Y}
\tpr_a(1-\tpr_a).
\end{aligned}
\]
} 
\blue{For PPV, however, when \(S \indep Y\), we have
\[
\ppv_a = \E(Y_i \mid D_i=1, A_i=a) = \E(Y_i \mid A_i=a) = \mu_a^Y.
\]
Thus the IF for the supervised, $(\mu_a^D)^{-1} D_i (Y_i - \ppv_a)$ and the IF for the Infairness, $(\mu_a^D)^{-1} D_i (Y_i - \mu_a^Y)$ are identical.
}

\clearpage 
\section{Real Data Analysis} 
\subsection{EHR-based phenotyping}\label{supp:real-data-training}
\blue{We construct the prediction using SuperLearner, an ensemble method that combines multiple ML algorithms,} with a silver-standard depression label using a random sample of 3,000 unlabeled observations. \blue{Specifically, the SuperLearner library includes regularized generalized linear models (SL.glmnet), Bayesian generalized linear models (SL.bayesglm), random forests (SL.ranger), gradient boosting machines (SL.xgboost), and a mean model as a baseline. } The silver-standard label deemed patients as having depression if (i) two depression-related Concept Unique Identifiers (CUIs) were present in the patient's discharge summary (C4049644 and C0011581) or (ii) there were depression-related International Classification of Diseases, Ninth Revision (ICD-9) codes (296.20 -- 296.26, 296.30 -- 296.36) in the patient's record. The features, $\bX$, included 22 other relevant CUIs extracted from discharge summaries using cTAKES software and 5 structured EHR features, including counts of evaluations, medicine notes, surgery notes, ICU stays, and prescriptions. The final classification for depression was obtained by thresholding the predictions at a cut-off that achieves overall FPR around 10\%.

\subsection{\blue{Chest X-ray audit}}\label{supp:chest}
\blue{We study classification of cardiomegaly from the CheXpert dataset, which contains 224{,}316 chest radiographs from 65{,}240 patients \citep{10.1609/aaai.v33i01.3301590}. We randomly selected one image per patient for fairness auditing. The labeled dataset consists of $n = 500$ radiologist-labeled images while the remaining $N = 64{,}740$ images are unlabeled. The prediction $S$ is obtained from a publicly available pre-trained DenseNet-121 model with weights from TorchXRayVision \citep{pmlr-v172-cohen22a}. We evaluate fairness across sex (male vs female;  55.5\% vs 44.5\%) and also have available an auxiliary covariate, age.}

\begin{figure}[htbp!]
    \centering
    \includegraphics[width=\linewidth]{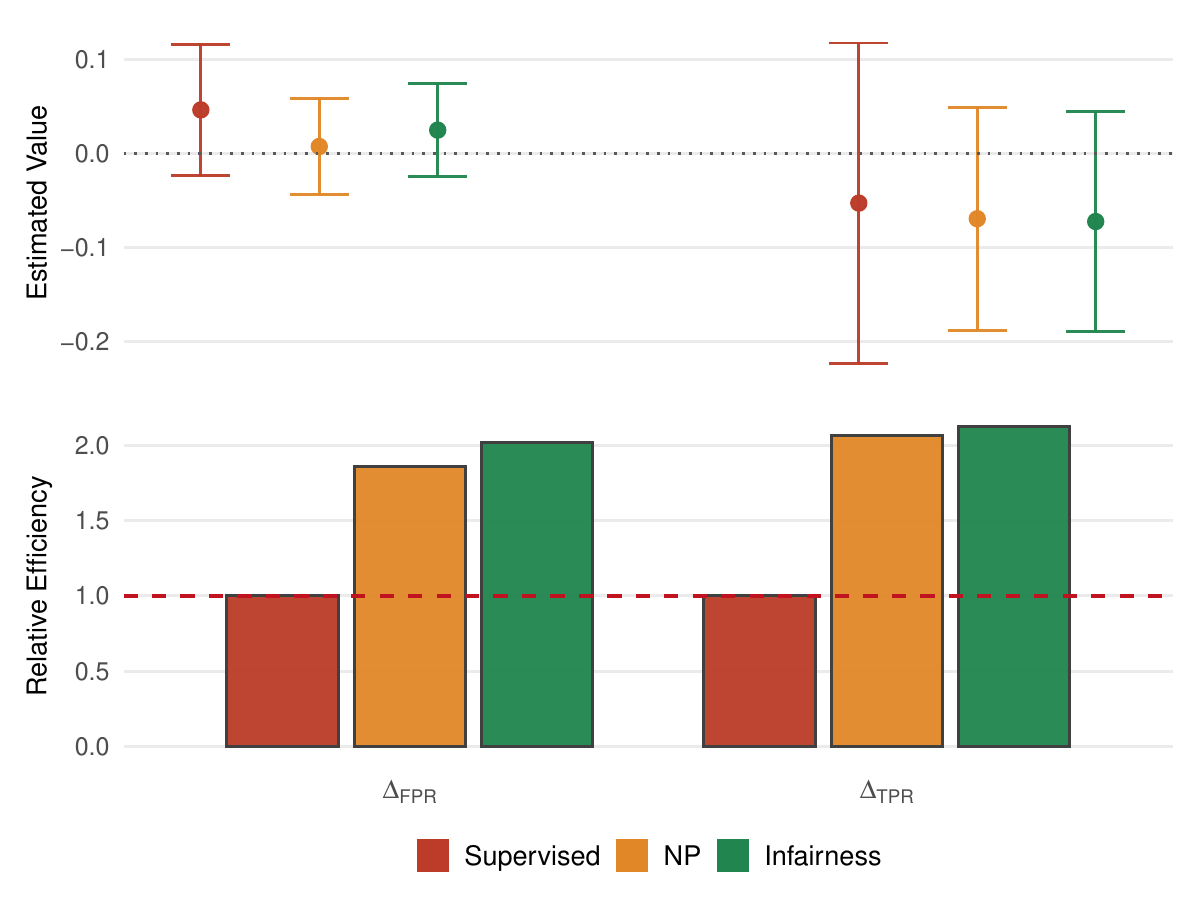}
    \caption{\blue{\textbf{FPR and TPR disparity estimates Chest X-ray audit across sex.} Disparities are defined as the difference in group-specific performance between the female and male. }}
    \label{fig:chest}
\end{figure}

\blue{Figure \ref{fig:chest} shows results for the audit across sex. Similar to the prior example, the supervised, NP, and Infairness estimators yield relatively similar point estimates. The 95\% confidence intervals for $\Delta_{\fpr}$ suggest that predictive equality is satisfied while Infairness suggests a potential violation of equal opportunity. The RE of Infairness is 2.56 for $\Delta_{\tpr}$ and 1.97 for $\Delta_{\fpr}$, which corresponds to variance reductions of 60.9\% and 56.3\%. The efficiency gains of Infairness relative to NP are less substantial than the EHR phenotyping example, likely due to the inclusion of a single auxiliary covariate. Prior work demonstrates that chest X-ray classifiers exhibit variable performance across socioeconomic status and race \citep{seyyed2021underdiagnosis}. These variables could yield higher precision of Infairness, but were unavailable in the version of our dataset.}

\clearpage
\vskip 0.2in

\putbib[references]

\end{bibunit}

\end{document}